\documentclass[onefignum,onetabnum]{siamonline190516}

\usepackage{amsmath, amssymb, mathrsfs}
\usepackage{physics}
\usepackage{lipsum}
\usepackage{amsfonts}
\usepackage{graphicx,mathtools}
\usepackage{algorithmic}
\usepackage{booktabs}
\usepackage[titletoc]{appendix}
\usepackage{bm}

\usepackage{threeparttable}
\usepackage{subcaption} 
\usepackage{siunitx}
\usepackage{multirow}
\usepackage{placeins}
\usepackage[export]{adjustbox}

\usepackage{enumitem}
\setlist[enumerate]{leftmargin=.5in}
\setlist[itemize]{leftmargin=.5in}

\newtheorem{Th}{Theorem}[section]
\newtheorem{Lemma}[Th]{Lemma}

\newtheorem{Cor}[Th]{Corollary}
\newtheorem{Def}[Th]{Definition}

\newtheorem{Rem}[Th]{Remark}

\newtheorem{Ass}[Th]{Assumption}

\def\R{{\mathbb R}}

\def\N{{\mathbb N}}

\def\E{{\mathbb E}}

\def\L{{\mathcal L}}

\def\P{{\mathbb P}}

\def\V{{\mathrm{Var}}}
\def\C{{C}}

\def\e{{\varepsilon}}

\def\v{{\mathbf{v}}}

\def\var{{\mathrm{Var}}}
\def\ex{{\mathsf{r}}}
\def\ext{{\mathsf{t}}}

\def\opt{{\mathrm{opt}}}

\def\mse{{\mathsf{MSE}}}
\def\Tr{{\mathrm{Tr}}}
\def\iid{\text{i.i.d.}}

\def\lrmc{{\mathsf{LRMC}_{\mathrm{opt}}}}
\def\lrmcstar{{\mathsf{LRMC}^*_{\mathrm{opt}}}}
\def\lrmco{{\mathsf{LRMC}}}

\def\bt{{\widehat{\beta}}}
\def\sg{{\widehat{\sigma}}}
\def\Cov{{\mathrm{Cov}}}
\def\blue{{\mathrm{BLUE}}}

\DeclareMathOperator*{\argmin}{arg\,min}

\def\lfs{{{[n]}}}
\def\afs{{{[n]_{+}}}}

\usepackage{xspace}
\usepackage{bold-extra}
\usepackage[most]{tcolorbox}

\colorlet{texcscolor}{blue!50!black}
\colorlet{texemcolor}{red!70!black}
\colorlet{texpreamble}{red!70!black}
\colorlet{codebackground}{black!25!white!25}

\lstdefinestyle{siamlatex}{%
  style=tcblatex,
  texcsstyle=*\color{texcscolor},
  texcsstyle=[2]\color{texemcolor},
  keywordstyle=[2]\color{texemcolor},
  moretexcs={cref,Cref,maketitle,mathcal,text,headers,email,url},
}

\tcbset{%
  colframe=black!75!white!75,
  coltitle=white,
  colback=codebackground, 
  colbacklower=white, 
  fonttitle=\bfseries,
  arc=0pt,outer arc=0pt,
  top=1pt,bottom=1pt,left=1mm,right=1mm,middle=1mm,boxsep=1mm,
  leftrule=0.3mm,rightrule=0.3mm,toprule=0.3mm,bottomrule=0.3mm,
  listing options={style=siamlatex}
}

\newtcblisting[use counter=example]{example}[2][]{%
  title={Example~\thetcbcounter: #2},#1}

\newtcbinputlisting[use counter=example]{\examplefile}[3][]{%
  title={Example~\thetcbcounter: #2},listing file={#3},#1}

\DeclareTotalTCBox{\code}{ v O{} }
{ 
  fontupper=\ttfamily\color{black},
  nobeforeafter,
  tcbox raise base,
  colback=codebackground,colframe=white,
  top=0pt,bottom=0pt,left=0mm,right=0mm,
  leftrule=0pt,rightrule=0pt,toprule=0mm,bottomrule=0mm,
  boxsep=0.5mm,
  #2}{#1}

\patchcmd\newpage{\vfil}{}{}{}
\begin{tcbverbatimwrite}{tmp_\jobname_header.tex}
\title{Optimally balancing exploration and exploitation to automate multi-fidelity statistical estimation\thanks{Submitted to the editors on \today.
}}

\author{
Thomas Dixon\thanks{Department of Aerospace Engineering, University of Michigan, Ann Arbor, MI, and Optimization and Uncertainty Quantification, Sandia National Laboratories, Albuquerque, NM.}
\and Alex Gorodetsky\thanks{Department of Aerospace Engineering, University of Michigan, Ann Arbor, MI.}
\and John Jakeman\thanks{Optimization and Uncertainty Quantification, Sandia National Laboratories, Albuquerque, NM.}
\and Akil Narayan\thanks{Scientific Computing and Imaging Institute, and Department of Mathematics, University of Utah, UT.}
\and Yiming Xu\thanks{Department of Mathematics, University of Kentucky, KY 
  (\email{yiming.xu@uky.edu}).}}

\headers{Automating multi-fidelity statistical estimation}{T. Dixon and A. Gorodetsky and J. Jakeman and A. Narayan and Y. Xu}
\end{tcbverbatimwrite}
\title{Optimally balancing exploration and exploitation to automate multi-fidelity statistical estimation\thanks{Submitted to the editors on \today.
}}

\author{
Thomas Dixon\thanks{Department of Aerospace Engineering, University of Michigan, Ann Arbor, MI, and Optimization and Uncertainty Quantification, Sandia National Laboratories, Albuquerque, NM.}
\and Alex Gorodetsky\thanks{Department of Aerospace Engineering, University of Michigan, Ann Arbor, MI.}
\and John Jakeman\thanks{Optimization and Uncertainty Quantification, Sandia National Laboratories, Albuquerque, NM.}
\and Akil Narayan\thanks{Scientific Computing and Imaging Institute, and Department of Mathematics, University of Utah, UT.}
\and Yiming Xu\thanks{Department of Mathematics, University of Kentucky, KY
  (\email{yiming.xu@uky.edu}).}}

\headers{Automating multi-fidelity statistical estimation}{T. Dixon and A. Gorodetsky and J. Jakeman and A. Narayan and Y. Xu}

\ifpdf
\hypersetup{pdftitle={Guide to Using  SIAM'S \LaTeX\ Style} }
\fi

\begin{document}

\maketitle

\begin{abstract}
Multi-fidelity methods that use an ensemble of models to compute a Monte Carlo estimator of the expectation of a high-fidelity model can significantly reduce computational costs compared to single-model approaches. These methods use oracle statistics, specifically the covariance between models, to optimally allocate samples to each model in the ensemble. However, in practice, the oracle statistics are estimated using additional model evaluations, whose computational cost and induced error are typically ignored. To address this issue, this paper proposes an adaptive algorithm to optimally balance the resources between oracle statistics estimation and final multi-fidelity estimator construction, leveraging ideas from multilevel best linear unbiased estimators in Schaden and Ullmann (2020) and a bandit-learning procedure in Xu et al. (2022). Under mild assumptions, we demonstrate that the multi-fidelity estimator produced by the proposed algorithm exhibits mean-squared error commensurate with that of the best linear unbiased estimator under the optimal allocation computed with oracle statistics. Our theoretical findings are supported by detailed numerical experiments, including a parametric elliptic PDE and an ice-sheet mass-change modeling problem.
\end{abstract}



\section{Introduction}

Estimation of the expectation of scalar-valued quantities of interest (QoIs) is a ubiquitous task in computational sciences that arises, for example, when quantifying uncertainty in model predictions. Most existing estimation methods are either based on Monte Carlo (MC) quadrature, Quasi-MC quadrature~\cite{sobol95}, or high-order quadrature such as cubature rules~\cite{Xiu_ANM_2008}, tensor product quadrature and sparse grid quadrature~\cite{Smolyak_SMD_1963, bungartz2004sparse}. In particular, when there is exploitable high-order structure in QoIs, adaptive collocation schemes are amongst the most efficient methods for such problems \cite{schwab_sparse_2011,piazzola_uncertainty_2020,narayan_adaptive_2014,liu_adaptive_2011}.  However, without such exploitable structure, accurately estimating the expectation of a computationally expensive model with these methods can be intractable. For example, high-order quadrature methods converge quickly when a small number of input parameterize the model, but typically require a number of samples that can grow rapidly, in some cases exponentially fast, with the number of inputs.  In contrast, the convergence rate of the error in MC and Quasi-MC quadrature estimators is independent or weakly dependent on the number of inputs, but their convergence rate is slow, and thus these methods can require a large number of samples to achieve even moderate accuracy. To overcome these challenges, this paper presents a practical adaptive multi-fidelity MC quadrature algorithm under a fixed computational budget, which uses multiple models of varying accuracy and cost (referred to as fidelities), to estimate expectations of high-dimensional functions at a fraction of the cost of single-fidelity MC quadrature, which uses a single model.

The accuracy of an MC estimator is typically measured by the mean-squared error (MSE). In the ideal case where potential bias from additional numerical errors is ignored, the MSE of a single-fidelity estimator of the expectation of a function is equal to the variance of the function divided by the number of samples used. Thus, the MSE of a single fidelity MC estimator can only be reduced by increasing the number of samples of a model, because the variance of a function is fixed. In contrast, multi-fidelity MC estimators can leverage the correlation between multiple models to further reduce the variance, leading to a smaller MSE of the estimated expectation of a high-fidelity model. Multi-fidelity methods such as multilevel Monte-Carlo (MLMC) estimators \cite{giles2008multilevel}, multi-fidelity Monte-Carlo (MFMC) estimators \cite{Peherstorfer_2016}, Approximate Control Variates (ACVs) \cite{Gorodetsky_2020}, and Multilevel best linear unbiased estimators (MLBLUEs) \cite{Schaden_2020}, have all been successfully used to reduce the computational cost of computing expectations of applications of computationally expensive models, in some cases by orders of magnitude. This increased performance is obtained by judiciously allocating the number of evaluations of models of varying fidelity that are highly correlated with the high-fidelity model but are less computationally expensive to simulate. Examples of such models include those derived from varying numerical discretizations and those based on idealized physics.   

Formally, multi-fidelity MC quadrature methods strategically allocate samples to an ensemble of models by determining the optimal allocation that minimizes the MSE of the multi-fidelity estimator. However, solving such optimization problems requires oracle cross-model statistics, such as the covariance between models, when estimating expectations. These statistics are typically estimated using a small number of model evaluations in a so-called \textit{pilot study}. The impact of the error in the estimated oracle statistics on the optimization problem is often ignored. Moreover, the computational cost of collecting these samples is often overlooked~\cite{jakeman_PSHHHP_Egusphere_2024}. A major focus of this paper is to optimally account for these factors in the construction of a multi-fidelity MC estimator, assuming a fixed computational budget.

Under a budget constraint, constructing a multi-fidelity estimator must balance the resources between the pilot study and the estimation for the high-fidelity model. With this motivation, Reference \cite{xu2022bandit} proposed balancing the cost of this two-step exploration-exploitation trade-off using a multi-armed bandit algorithm~\cite{lattimore2020bandit}. Their so-called adaptive explore-then-commit algorithm (AETC) optimized the exploration budget assigned to computing oracle statistics with the exploitation cost of computing the expectation of the desired QoI. Moreover, while iteratively refining estimates of the oracle statistics, the AETC algorithm was able to choose the best set of models to produce a multi-fidelity estimator called the linear regression Monte-Carlo (LRMC) estimator that was substantially more efficient than the corresponding single-fidelity MC estimator constructed with the same computational budget. We briefly mention that there is additional work in the multi-fidelity community for balancing cost/effort in exploration and exploitation phases \cite{peherstorfer_multifidelity_2019,alsup_context-aware_2023}. These approaches make \textit{hierarchical} model assumptions and implicitly presume the ability to inform the construction of low-fidelity models. Moreover, they do not seek to learn cost-accuracy tradeoffs or correlations of a given and fixed ensemble of models, which is different from the AETC approach.

While identifying the promise of formulating multi-fidelity estimation as an exploration-exploitation trade-off, the study in \cite{xu2022bandit} had a limited scope. Specifically, the AETC algorithm adopted a uniform exploitation policy to construct an LRMC estimator by allocating the same number of samples to each selected low-fidelity model, resulting in a sub-optimal MSE in the LRMC. Indeed, our numerical results reveal a gap between the MSE of the LRMC estimator produced by AETC and a theoretical lower bound computed using oracle statistics. The exploration policy is also uniform, but optimizing over this step yields only limited improvement on overall accuracy under some common multi-fidelity assumptions (see \Cref{ajan}). 

To fill the gap, this article generalizes the AETC algorithm in \cite{xu2022bandit} by proposing a strategy to improve the LRMC estimator for the high-fidelity model. The key observation is that the MC estimator for the low-fidelity means in the LRMC estimator in \cite{xu2022bandit} can be replaced with a larger class of estimators while admitting a similar exploration-exploitation trade-off. This enables the improvement of the estimation efficiency while ensuring automated sample allocation. To this end, we consider an optimal estimator for the low-fidelity means based on the ideas of MLBLUEs and show that it falls within the extended framework. Under these circumstances, we analyze the corresponding algorithm (AETC-OPT in \Cref{alg2}) and provide theoretical guarantees for the resulting estimator $\lrmc$ in terms of consistency, optimality, and robustness. Part of these results is established by connecting $\lrmc$ to other well-studied multi-fidelity estimators. We also conduct numerical experiments to demonstrate that our estimators have similar MSE to the MLBLUE under the optimal allocation computed with oracle information (even without accounting for oracle estimation costs). \Cref{tab:1} provides a brief comparison between this work and a subset of the existing literature. 

\begin{table}[h!]
\centering
\renewcommand{\arraystretch}{1.3}
\setlength{\tabcolsep}{8pt}
{\small
\begin{tabular}{@{}lll@{}}
\toprule
\textbf{Estimators} & \textbf{Automated Sample Allocation} & \textbf{Optimal Allocation} \\ 
\midrule
ACVs \cite{Gorodetsky_2020} / MLBLUEs \cite{Schaden_2020} & No  & --- \\
$\lrmco$ \cite{xu2022bandit}                                  & Yes (AETC algorithm)     & No \\
$\lrmc$ (this work)                        & Yes (AETC-OPT algorithm) & Yes \\
\bottomrule
\end{tabular}
}
\caption{Comparison of $\lrmc$ with other multi-fidelity estimators in the literature. 
For methods that do not require pilot studies, we list the sequential algorithms used to determine optimal sample allocation under a given budget.}
\label{tab:1}
\end{table}

The remainder of the paper is organized as follows. \Cref{sec:background} reviews the background, including best linear unbiased estimators (\Cref{sec:BLUE}), a direct approach for incorporating pilot study costs in the sample allocation optimization (\Cref{sec:problem}), and a sequential approach presented in previous work on LRMC/AETC (\Cref{sec:previouswork}). \Cref{sec:efficient} generalizes the AETC algorithm by introducing more efficient exploitation estimators. Specifically, \Cref{sec:new_MSE} develops the theory for generalization, and \Cref{sec:algorithm} introduces the AETC-OPT algorithm, which adaptively selects the number of pilot samples for the improved LRMC estimator $\lrmc$. \Cref{sec:exploitationBLUE} analyzes AETC-OPT's asymptotic properties, robustness, and optimality versus a theoretical lower bound, while \Cref{sec:numerical} presents an extensive numerical study. All proofs and additional numerical details are provided in the Appendices.

\section{Background}\label{sec:background}

We present a method for optimizing the exploration-exploitation trade-off when constructing multi-fidelity estimators of expectations. While much of the analysis applies to general multi-fidelity estimators, in this article, we will focus on the use of MLBLUEs; see~\cite{Gorodetsky_2024} for an exposition on the relationships between MLBLUEs and other linear multi-fidelity estimators. Consequently, in this section, we review the construction of MLBLUEs and highlight the challenge of directly optimizing the exploration-exploitation trade-off.

\subsection{MLBLUEs} \label{sec:BLUE}
Let $\lfs = \{1, \ldots, n\}$ and $\afs = \{0\}\cup [n]$. Given a total budget $B$ (e.g., afforded simulation time on a computer), an MLBLUE can be used to estimate the mean of the random scalar output, $Q_0$, of a high-fidelity model, with random cost $\C_0$ with mean $c_0$, using the random scalar outputs of $n$ low-fidelity models represented by $\{Q_i\}_{i\in\lfs}$ with random costs $\{\C_i\}_{i\in\lfs}$ with means $\{c_i\}_{i\in\lfs}$. The expectations of $\{Q_i\}_{i\in\afs}$ are denoted by $\{\mu_i\}_{i\in\afs}$, respectively. We also assume $\{\C_i\}_{i\in\afs}$ are uniformly bounded a.s. The random cost setup is of potential interest for applications where mean costs are not available but can be sequentially estimated. For the theoretical analysis in the remainder of the paper, we assume the costs are deterministic and equal to their means unless otherwise stated (e.g., in \Cref{sec:algorithm}).

Denote the powerset of $\afs$ as $\mathcal{P}(\afs)$. An MLBLUE consists of a linear combination of correlated MC estimators constructed using subsets $T\in\mathcal{P}(\afs)$ of available model fidelities. Letting $Q_T = (Q_i)_{i\in T}$ represent a $|T|$-dimensional vector of the scalar outputs of the subset $T$ of low-fidelity models, each MC estimator is built by drawing a set of independent samples from $Q_T$. We denote $W_{T,\ell}$ as the $\ell$th sample of the model set $T$. Typically, realizations of $Q_T$ are generated by drawing independent samples from the joint distribution of the model inputs and then evaluating each model in $T$ at those input samples. Thus, the cost of building each estimator is determined by the cost of evaluating all models in $T$ at a given sample and the number $m_T$ of independent samples of $Q_T$. A fixed number of samples of each model is encoded in a \textit{sample allocation} $\mathcal{M}: \mathcal{P}([n]_+) \rightarrow \N$, with $m_T \coloneqq \mathcal{M}(T)$. The expected cost of constructing an MLBLUE is $\sum_{T \subseteq \afs}c_T m_T$. 

Denoting the mean vector of all models as $\mu_\afs = (\mu_0, \ldots, \mu_n)^\top$, the MLBLUE for $\mu_\afs$ under sample allocation $\mathcal M$ can be derived from the following linear model formulation
\begin{align}
\begin{bmatrix}
\vdots\\
W_{T,1}\\
\vdots\\
W_{T,m_T}\\
\vdots
\end{bmatrix} = 
\begin{bmatrix}
\vdots\\
R_{T}\\
\vdots\\
R_{T}\\
\vdots
\end{bmatrix}
\mu_\afs +
\begin{bmatrix}
\vdots\\
\xi_{T, 1}\\
\vdots\\
\xi_{T, m_T}\\
\vdots
\end{bmatrix}\label{2}
\end{align}
that uses the restriction matrices $R_T\in\{0, 1\}^{|T|\times (n+1)}$ to map $\mu_\afs \in\R^{n+1}$ to a vector $\mu_T\in\R^{|T|}$ with component indices in $T$. The heterogeneous noise vector is formed by concatenating independent centered random vectors; block $T$ of this concatenated vector has entries $\xi_{T,\ell} = W_{T,\ell}-\mu_T$, each with covariance $\Sigma_T = \E[(Q_T-\mu_T)(Q_T-\mu_T)^\top]$.  Using more compact notation, we rewrite \eqref{2} as
\begin{align}
    W = R \mu_\afs + \xi \label{eq:allBLUE}
\end{align}
where $W= (W_{T, \ell})$, $R= (R_{T})$, and $\xi= (\xi_{T, \ell})$ for all $\ell\in[m_T]$ and $T\in\mathcal{P}(\afs)$.

Let $\Sigma = \Cov[\xi]$ (i.e., $\Sigma\neq \Sigma_{\afs}$ in general). The Gauss--Markov theorem \cite{johnson2014applied} states that, among all linear unbiased estimators for $\mu_\afs$, where linearity is with respect to $W$, the MLBLUE, defined as a reweighted least-squares solution to \eqref{eq:allBLUE},
\begin{align}
\widehat{\mu}_{\blue} &:= (R^\top \Sigma^{-1} R)^{-1} R^\top \Sigma^{-1} W \nonumber\\
&= \left(\sum_{T\subseteq \afs }m_TR_T^\top \Sigma_T^{-1}R_T\right)^{-1}\sum_{T\subseteq \afs}R_T^\top \Sigma_T^{-1}\sum_{\ell\in [m_T]}W_{T,\ell},\label{eq:BLUE_estimator}
\end{align}
has the Loewner-order minimum covariance. This minimum covariance, that of $\widehat{\mu}_\blue$, is 
\begin{align*}
\Cov[\widehat{\mu}_{\blue}] = \left(\sum_{T\subseteq \afs}m_TR_T^\top \Sigma_T^{-1}R_T\right)^{-1}.
\end{align*}
Consequently, for any $a\in\R^{n+1}$,
\begin{align}
\V[a^\top \widehat{\mu}_{\blue}] & = a^\top \left(\sum_{T\subseteq \afs}m_TR_T^\top \Sigma_T^{-1}R_T\right)^{-1}a\label{optf}
\end{align}
is the smallest variance of the $a$-sketch of $\mu_\afs$ achievable by all linear unbiased estimators under the fixed allocation $\mathcal M$. Thus, the optimal allocation that minimizes the variance of this $a$-sketch subject to the cost of the estimator being less than the budget $B$ can be found by solving the optimization problem
\begin{align}
  \min_{\mathcal M = \{m_T\}_{T\subseteq \afs}} \V[a^\top \widehat{\mu}_{\blue}]\hskip 10pt \textrm{subject to } \sum_{T\subseteq \afs}c_T m_T\leq B,\label{optint}
\end{align}
where we identify a sample allocation $\mathcal M$ as a vector $\{m_T\}_{T\subseteq \afs}$. 
Typically, a relaxed version of this optimization problem, which optimizes continuous values of the number of samples rather than integers, is used to find an optimal allocation. The final allocation is then found by rounding each $m_T$ down to the nearest integer. In this paper, we solve this relaxed optimization problem using semi-definite programming as proposed in~\cite{Croci_WW_CMAME_2023}.

\subsection{A Direct Approach}\label{sec:problem}

The variance of an MLBLUE \eqref{optf} depends on the covariance of the model outputs, $\Sigma_\afs$, which is often not known \textit{a priori}. Therefore, these statistics need to be estimated using a set of pilot samples before determining the optimal sample allocation. However, the computational cost and the error introduced by estimating these oracle statistics are often overlooked. To our knowledge, most existing multi-fidelity estimators, including alternatives to MLBLUEs, suffer from this limitation. An alternative to conducting a pilot study is ensemble estimation \cite{Pham_G_SIAMUQ_2022}, which fixes a total budget and estimates the necessary oracle statistics using an ensemble of estimators. This approach considers the cost, but continues to ignore the error introduced by incorrect statistics.
 
More recently, \cite{dixon2024covariance} demonstrated the decaying performance of ACV estimation as the percentage of the budget given to a pilot study is varied. This suggests that the cost of a pilot study can substantially impact the accuracy of an MLBLUE; see our numerical results in \Cref{sec:ice-sheet}. The exact cost of the pilot study is the computational cost of evaluating $q$ independent draws of the model outputs, which are used to compute the oracle statistics needed, e.g., the covariance between models when using an MLBLUE. Consequently, the average cost of the pilot is given by $q c_\afs$, where we note that $c_\afs$ is the average cost of evaluating all models once. Thus, the total average cost of building an MLBLUE is 
\begin{align}
\mathcal{C}(q,\mathcal{M})=q c_\afs+\sum_{T\subseteq \afs}c_T m_T. \label{eq:total-est-ost} 
\end{align}

Existing methods ignore the cost of the pilot study, i.e., the first term in \eqref{eq:total-est-ost}, when optimizing the sample allocation of a model. However, this cost involves evaluating the high-fidelity model, which is computationally expensive and should be accounted for from an economic perspective. A direct approach to incorporate this additional cost is to consider the following optimization problem:
\begin{align}
  \min_{q,\mathcal{M}} \quad \E[(\widehat{\mu}_0(q,\mathcal{M})- \mu_0)^2], \hskip 10pt \textrm{subject to } \mathcal{C}(q,\mathcal{M}) \leq B,  \label{eq:MSE_min}
\end{align}
where $\widehat{\mu}_0(q,\mathcal{M})$ is the approximate MLBLUE for $\mu_0$ under sample allocation $\mathcal M$ and estimated covariance $\widehat{\Sigma}$ based on $q$ pilot samples. However, solving \eqref{eq:MSE_min} is rather difficult as it requires deriving an implementable expression for the MSE. To see this, let $e_0 = (1,0,\ldots,0)^\top\in \R^{n+1}$ so that
\begin{align}
\widehat{\mu}_0(q,\mathcal{M}) = e_0^\top \Psi W\quad\quad \Psi \equiv (R^\top \widehat{\Sigma}^{-1} R)^{-1} R^\top \widehat{\Sigma}^{-1}.
\end{align}
If the $q$ pilot samples are independent from the sample allocation $\mathcal{M}$, then $\widehat{\mu}_0(q,\mathcal{M})$ is unbiased for $\mu_\afs$ since
\begin{align}
\E[\widehat{\mu}_0(q,\mathcal{M})] = e_0^\top\E[\Psi W] = e_0^\top\E[\Psi] \E[W] = e_0^\top \mu_\afs = \mu_0.\label{appunb}
\end{align}
As a result, the MSE for $\widehat{\mu}_0(q,\mathcal{M})$ reduces to the variance: 
\begin{align}
  &\E[(\widehat{\mu}_0(q,\mathcal{M}) - \mu_0)^2] = \V[\widehat{\mu}_0(q,\mathcal{M})]  = \V[e_0^\top\Psi W] \nonumber\\
    =&\ \Tr( \Cov[e_0^\top\Psi]\Cov[W] ) + \E[W]^\top\Cov[e_0^\top\Psi]\E[W] + \E[e_0^\top\Psi]^\top\Cov[W]\E[e_0^\top\Psi],
    \label{eq:BLUE_MSE}
\end{align}
where the last step follows from a direct calculation using independence (see \Cref{ap:independent_variance} for a detailed derivation). 

The objective in \eqref{eq:BLUE_MSE} involves complicated terms such as $\E[\Psi]$ and $\V[\Psi]$ and thus is intractable to handle in general. In the subsequent sections, we present an alternative bandit-learning approach to account for pilot study costs.

\subsection{LRMC/AETC}\label{sec:previouswork}

Because finding the MSE of an optimized multi-fidelity estimator, such as an MLBLUE, is challenging when considering pilot samples, Reference \cite{xu2022bandit} introduced LRMC estimators as an alternative for optimization in a model-assisted framework. Particularly, the MSE of an LRMC estimator can be decomposed into two error terms associated with the pilot sampling (exploration) and estimator construction (exploitation) phases. This decomposition is more computationally tractable and can be used for adaptive sample allocation within a multi-armed bandit algorithm, with arms identified as model subsets $T$. From here on, pilot samples will now be denoted as exploration samples for clarity.

The LRMC estimators were motivated by a joint linear model assumption between the high-fidelity and low-fidelity models, although the estimators themselves are applicable more broadly. For every $S\subseteq [n]$ with $|S|=s$, an LRMC estimator associated with $S$ is constructed by assuming that the high and low-fidelity models with indices in $S$, $Q_0$ and $Q_S$ respectively, satisfy the linear model:
\begin{align}
&Q_0 = Q_S^\top b_S + a_S + \e_S&\e\sim (0, \sigma_S^2),\label{lr}
\end{align}
where $b_S\in\R^{s}$ is the model coefficient vector, $a_S\in\R$ is the intercept, and $\e_S$ is the discrepancy term with conditional expectation and variance $\E[\e_S | Q_S] = 0$ and $\E[\e_S^2 | Q_S] = \sigma_S^2$, respectively. Assuming $a_S$ and $b_S$ are known, we can take the expectation of both sides of \eqref{lr} to obtain an estimator for $\mu_0$ by estimating the low-fidelity means $\mu_S$:
\begin{align}
    \widehat{\mu}_0 = \widehat{\mu}_S^\top b_S + a_S.\label{eq:AETC_format}
\end{align}

In contrast to the global optimization problem in \eqref{eq:MSE_min}, the added flexibility of choosing $S$ to form a linear model in \eqref{lr} can be interpreted as dissecting \eqref{eq:MSE_min} into a number of competing subproblems (for optimization) by imposing a sparsity constraint on the allocation vector $\mathcal M = \{m_T\}_{T\subseteq\afs}$. Restricting to the low-fidelity index set $S$ is equivalent to requiring $m_T = 0$ for $T$ such that $T \cap (S \cup \{0\})^\complement \neq \emptyset$. 

In practice, both $a_S$ and $b_S$ are unknown and must be estimated from the data. Reference \cite{xu2022bandit} proposed an AETC multi-armed bandit procedure to balance the exploration cost of computing the coefficients, $a_S$ and $b_S$, and the exploitation cost of computing the low-fidelity means. Their proposed algorithm employed a uniform exploration strategy that evaluated all models on the same set of samples to estimate $a_S$ and $b_S$, followed by an exploitation phase that utilized the remaining budget to estimate the selected low-fidelity means. The LRMC estimator was optimized by finding the number of exploration and exploitation samples that minimized a derived expression for the MSE of an LRMC estimator conditioned on the exploration samples. 

To describe the ingredients in the AETC algorithm, we fix a total budget $B$ and simplify notation as follows: 
\begin{align}
Y_S =  (1, Q^\top_S)^\top\in\R^{s+1} \quad\quad\beta_S = (a_S, b^\top_S)^\top\in\R^{s+1} \quad\quad\textrm{such that}\quad  Q_0 = Y^\top_S \beta_S + \e_S. \label{1432!}
\end{align}
For a fixed $q$, the AETC algorithm comprises the following two steps.

\paragraph{Exploration}
Collect $q$ joint samples of every available model $(Q_0, \ldots, Q_n)$ for exploration. Let $c_{\ex} = c_\afs = \sum_{i=0}^n c_i$ denote the cost per exploration sample, and  
\begin{align}
&Z_S = (Y_{S,1}, \ldots, Y_{S, q})^\top\in\R^{q\times (s+1)} & Z_0 = (Q_{0,1}, \ldots, Q_{0,q})^\top \in\R^{q},\label{mydesign}
\end{align}
where $Y_{S,\ell} = (1, Q^\top_{S,\ell})^\top\in\R^{s+1}$ and $Q_{S,\ell} = (Q_{\ell, i})_{i\in S}^\top\in\R^s$ correspond to the restriction of the $\ell$th exploration sample restricted to index $S$. We can estimate $\beta_S$ using least squares: 
\begin{align}
\bt _S= Z_S^\dagger Z_0 = (Z_S^\top Z_S)^{-1}Z_S^\top Z_0,\label{lse}
\end{align}
where $Z_S$ is assumed to have full column rank.

\paragraph{Exploitation} 
Use the remaining budget $B_\ext = B - c_\ex q$ to construct an LRMC for $\mu_0$ associated with some $S$ selected in the exploration phase: 
\begin{align}
&\lrmco = \frac{1}{m}\sum_{\ell\in [m]}V^\top_{S,\ell}\widehat{\beta}_S, \quad\quad\quad\quad V_{S,\ell} = (1, W_{S, \ell})^\top\in\R^{s+1}, \quad\quad\quad\quad W_{S,\ell}\stackrel{\iid}{\sim}  Q_S,\label{lrmc}
\end{align} 
where $m = \left\lfloor B_\ext/c_S\right\rfloor$ such that $\lfloor \cdot \rfloor$ is the floor operator and $c_S = \sum_{i\in S} c_i$.

The value of $q$ significantly impacts the MSE of the LRMC estimator in \eqref{lrmc}. Ideally, $q$ should be chosen to balance the contribution of all the errors introduced in the exploration and exploitation phase. However, as with MLBLUEs, deriving such an exact expression for LRMC estimators is challenging. Fortunately, under the joint linear regression assumption~\eqref{lr}, it is possible to obtain an asymptotic expression for the MSE of \eqref{lrmc} conditioned on the exploration design matrix $Z_\lfs$ \cite{xu2022bandit}: For $0<q<B/c_\ex$, 
\begin{align}
\E\left[(\lrmco-\mu_0)^2|~Z_\lfs\right] &\simeq \frac{\tr(y_S y_S^\top\Pi_S^{-1})\sigma_S^2}{q} + \frac{c_S b_S^\top\Sigma_S b_S}{B-c_\ex q}= \frac{\sigma_S^2}{q} + \frac{c_S b_S^\top\Sigma_S b_S}{B-c_\ex q}&a.s.,\label{Ls}
\end{align}
where 
\begin{align}
y_S = \E[Y_S]\quad\quad \Sigma_S = \Cov[Q_S]\quad\quad \Pi_S = \E[Y_S Y_S^\top].\label{snota}
\end{align} 
The second equality in \eqref{Ls} is not in \cite{xu2022bandit} but follows from a direct computation using Schur complements:
\begin{align}
y_S y_S^\top\Pi_S^{-1} &= \begin{bmatrix}
1 & \mu_S^\top\\
\mu_S& \mu_S\mu_S^\top
\end{bmatrix}
\begin{bmatrix}
1 & \mu_S^\top\\
\mu_S& \E[Q_SQ_S^\top]
\end{bmatrix}^{-1} \nonumber\\
&= \begin{bmatrix}
1 & \mu_S^\top\\
\mu_S& \mu_S\mu_S^\top
\end{bmatrix}
\begin{bmatrix}
1+\mu_S^\top\Sigma_S^{-1}\mu_S & -\mu^\top_S\Sigma_S^{-1}\\
-\Sigma_S^{-1}\mu_S& \Sigma_S^{-1}\\
\end{bmatrix}\nonumber\\
& = \mathrm{diag}(1, 0, \ldots, 0).\label{newobs}
\end{align}
The symbol $\simeq$ denotes asymptotic equality such that 
\begin{align}
a \simeq b \hskip 10pt \Longleftrightarrow \hskip 10pt \lim_{\min\{q, B-c_\ex q\}\to\infty}\frac{a(q)}{b(q)} = 1. \label{here}
\end{align}
The asymptotic \eqref{Ls} was then minimized with respect to $q$ to find the optimal number of exploration samples if exploiting $\lrmco$ in \eqref{lrmc}. Since the expression in \eqref{Ls} depends only on statistics that are estimable from exploration samples, an adaptive algorithm was proposed that increased the number of exploration samples $q$ until the sum of two terms in the MSE was minimized under the best $S$, which is also selected by AETC.

While $\lrmco$ in \eqref{lrmc} admits an explicit computation of the asymptotic conditional MSE in \eqref{Ls}, the estimator itself is suboptimal because the exploitation phase in \eqref{lrmc} samples each selected low-fidelity model uniformly when estimating $\mu_S$. Estimators like ACVs or MLBLUEs use optimal sample allocations across model fidelities to minimize their MSEs. In the next section, we generalize \eqref{Ls} by considering a class of multi-fidelity estimators for $\mu_S$, which leads to a broader class of LRMC estimators beyond \eqref{lrmc}, meanwhile admitting a similar asymptotic MSE decomposition as \eqref{Ls}.


\section{Efficient Exploitation}\label{sec:efficient}

The AETC algorithm in \cite{xu2022bandit} was, to our knowledge, the first attempt to quantify the exploration-exploitation trade-off that was overlooked in other multi-fidelity studies. Nevertheless, this algorithm has a suboptimal exploitation procedure, leaving room for improvement. In particular, the uniform exploitation phase employed, which uses MC estimators of each low-fidelity model with the same number of samples as in \eqref{lrmc}, can be improved upon by using a multi-fidelity estimator that allocates different numbers of samples to each model under the same exploitation budget. 

This section proposes replacing the MC estimator for $\mu_{S}$ in \eqref{lrmc} with a general class of estimators that satisfy exploration-unbiasedness, and whose corresponding LRMC estimator exhibits an asymptotic inverse-linear scaling property in the exploitation budget (abbreviated as the asymptotic scaling property). These properties ensure that a similar bandit-learning framework can be applied to adaptively automate sample allocation. Among the many such estimators, we focus on MLBLUE and consider the corresponding AETC analogue, AETC-OPT, where the ``OPT'' refers to an optimal exploitation phase. Naturally, a change of the exploitation-phase estimator necessitates retooling the asymptotics in \eqref{Ls}.

\subsection{A Generalized MSE Estimate}\label{sec:new_MSE}
Designing a multi-armed bandit algorithm requires a loss function that can be minimized during the multi-armed learning procedure. This section introduces the MSE of an extended AETC procedure with a general, non-adaptive exploitation estimator; an adaptive variant with respect to a special exploitation estimator is considered in \Cref{sec:algorithm}. Specifically, we present an MSE expression for the class of estimators for $\mu_{S}$ used in exploitation that are unbiased conditional on exploration, a property we refer to as exploration-unbiasedness. 

\begin{Def}[Exploration-unbiased exploitation]\label{def:exploration-unbiased}
Let $Z_S$ and $\bt_S$ be the design matrix and least-squares estimator in \eqref{mydesign} and \eqref{lse}, respectively. An estimator $\widehat{\mu}_S$ for $\mu_S$ is called \textit{exploration-unbiased} if $\E[\widehat{\mu}_S\mid Z_\lfs, \bt_S] = \mu_S$. 
\end{Def}

By definition, any unbiased estimator for $\mu_S$ that depends on exploitation samples only (thus independent of $Z_\lfs$ and $\bt_S$) is exploration-unbiased. However, as we will see later, there exist estimators that depend on the exploration data yet remain exploration-unbiased. The following lemma formalizes the impact of using an exploration-unbiased estimator to construct an LRMC estimator for exploitation on the bias-variance decomposition of the MSE.

\begin{Lemma}\label{lemma:bl-asymptotic-mse}
Let $\widehat{\mu}_S$ be an exploration-unbiased estimator for $\mu_S$ constructed using exploration and exploitation samples. Let $\widehat{y}_S = (1,\widehat{\mu}^\top_S)^\top$ and $\bt_S = (\widehat{a}_S,\widehat{b}^\top_S)^\top$, where $\bt_S$ is the least-squares estimator defined in \eqref{lse} with $\widehat{a}_S, \widehat{b}_S$ denoting the estimated intercept and low-fidelity coefficients, respectively. Assume that the linear model assumption \eqref{lr} holds and denote by $\eta_S = (\e_{S, 1}, \ldots, \e_{S, q})^\top\in\R^q$ the model discrepancy vector in the exploration phase. Suppose that the following asymptotic scaling property holds: 
\begin{align}
&\gamma(S) \coloneqq \lim_{\min\{q, B_\ext\}\to\infty}B_\ext\cdot \E\left[\widehat{b}_S ^\top\Cov\left[\widehat{\mu}_S \mid \widehat{b}_S, Z_\lfs\right]\widehat{b}_S \mid Z_\lfs\right] \in (0, \infty)&a.s., \label{gamma}
\end{align}
where $Z_{\lfs}$ is the exploration design matrix defined in \eqref{mydesign} and $B_\ext = B - c_\ex q$. The conditional MSE of the associated LRMC estimator ${\widehat{y}_S}^\top\widehat{\beta}_S$ on $Z_{\lfs}$ satisfies
\begin{align}
&\mse_S := \E[({\widehat{y}_S}^\top\widehat{\beta}_S - \mu_0)^2 \mid Z_\lfs] ~\simeq~ \frac{\sigma_S^2}{q} + \frac{\gamma(S)}{B-c_\ex q}, \label{asympform}
\end{align}
where $\sigma_S^2$ is defined in \eqref{lr} and $\simeq$ is interpreted in the sense of \eqref{here}. 
\end{Lemma}
The proof of \Cref{lemma:bl-asymptotic-mse} is in \Cref{sec:supp2}. The main difference between \eqref{asympform} and \eqref{Ls} is the form of the numerator in the second term (concerning the exploitation phase). To present the multi-armed bandit procedure associated with \eqref{asympform}, we will use \Cref{lemma:bl-asymptotic-mse} to define some important quantities. For any $S\subseteq [n]$, using \eqref{asympform} let 
\begin{align}
&\L_S (q) = \frac{k(S) }{q} + \frac{\gamma(S)}{B-c_\ex q},&&\text{where}&&k(S)  =\sigma_S ^2,\label{thisn}
\end{align}
which is a strictly convex function in its domain and attains a unique minimizer. In particular, the asymptotically best $q$ is
\begin{align}
q_S^*  = \argmin_{0<q<B/c_{\ex}}\L_S (q) = \frac{B}{c_{\ex}+\sqrt{\frac{c_{\ex}\gamma(S) }{k(S) }}},\label{pj1}
\end{align} 
with optimum  
\begin{align}
\L_S ^* &=  \frac{(\sqrt{c_{\ex}k(S) }+\sqrt{\gamma(S) })^2}{B}\propto \left(\sqrt{c_{\ex}k(S) }+\sqrt{\gamma(S) }\right)^2.\label{xiaodaihua}
\end{align}

To mimic the MLBLUE for the high-fidelity model mean under the optimal allocation over the full index set $[n]$, we also perform a model selection procedure that optimally chooses the subset of models that minimizes the estimator's minimum MSE in \eqref{xiaodaihua}. By choosing a subset $S\subseteq \lfs$ of low-fidelity models, we can choose the subset that eliminates unnecessary models for optimal performance. The optimal subset, $S$, is the one that has the smallest $\L_S ^*$ under the optimal exploration rate $q^*_S$: 
\begin{align}
S^* = \argmin_{S\subseteq [n]} \L_S ^* = \argmin_{S\subseteq [n]} \left(\sqrt{c_{\ex}k(S)}+\sqrt{\gamma(S)}\right)^2,\label{pj2}
\end{align}
where the right-hand side is assumed to have a unique minimizer. This assumption is not essential but helps simplify the results in the subsequent discussion.

\subsection{Algorithm}\label{sec:algorithm}
In this section, we focus on a special optimal choice of $\widehat{\mu}_S$ that satisfies exploration-unbiasedness, and whose corresponding LRMC estimator exhibits the asymptotic scaling property in \Cref{lemma:bl-asymptotic-mse}. Specifically, we take $\widehat{\mu}_S$ to be the MLBLUE in \eqref{eq:BLUE_estimator}, with the allocation set $\mathcal{M} = \{m_T\}_{T \subseteq S}$ chosen as the optimal allocation determined using the estimated sketch $\widehat{b}_S$ under the exploitation budget $B_\ext$ as in \eqref{optint}. 

If we further assume that $\gamma(S) \in (0, \infty)$ exists and is consistently estimable using the exploration data, then we can convert the discussion in \Cref{sec:new_MSE} into an adaptive algorithm. The technical properties of $\widehat{\mu}_S$ and the estimability of $\gamma(S)$ will be addressed in \Cref{sec:exploitationBLUE}. In the following, we denote by $\widehat{\gamma}_q(S)$ its estimator based on the first $q$ exploration samples. To accommodate the scenario of potentially random costs, we treat sampling costs as random in the algorithm description.

We now construct an adaptive explore-then-commit algorithm using empirical estimators. Specifically, for $S\subseteq [n]$ and exploration samples $\{Q_{\afs,\ell}\}_{\ell\in [q]}$, and costs $\{\C_{\lfs, \ell}\}_{\ell\in [q]}$ ($q\geq n+2$), we define
\begin{align}
& \widehat{c}_S = \frac{1}{q}\sum_{\ell\in [q]}\C_{S, \ell}& \bar{y}_{S}& = (1, \Bar{\mu}^\top_S)^\top = \frac{1}{q}\sum_{\ell\in [q]}Y_{S,\ell}\nonumber\\
&\sg_S^2 = \frac{1}{q-|S|-1}\sum_{\ell\in [q]}\left(Q_{0,\ell}-Y_{S,\ell}^\top\bt_S\right)^2& \widehat{k}_q(S)& = \widehat{\sigma}_S^2
\label{haozi}
\end{align}
where $\widehat{\beta}_S$ is defined in \eqref{lrmc} and we used $\bar{y}_{S}$ to differentiate from the exploitation estimator $\widehat{y}_S$.  
Thus, the asymptotic loss function $\L_S$, its minimizer, and optimum can be estimated as 
\begin{align}
\widehat{\L}_S(z; q) &= \frac{\widehat{k}_q(S)}{z} + \frac{\widehat{\gamma}_q(S)}{B-\widehat{c}_\ex z}\label{eq:estimated-loss}\\
\widehat{q}^*_S(q) &= \argmin_{0<z<B/\widehat{c}_{\ex}}\widehat{\L}_S(z; q) = \frac{B}{\widehat{c}_{\ex}+\sqrt{\frac{\widehat{c}_{\ex}\widehat{\gamma}_q(S)}{\widehat{k}_q(S)}}}&\widehat{\L}_S^*(q) &=  \frac{(\sqrt{\widehat{c}_{\ex}\widehat{k}_q(S)}+\sqrt{\widehat{\gamma}_q(S)})^2}{B}.\label{getq*}
\end{align}
Note that above we have two exploration sample counters: $q$ denotes the number of samples used to construct estimators (i.e., samples already collected), and $z$ denotes a putative number of eventual exploration samples collected for tentative decision-making. The estimated optimal loss of exploiting models $S$ after $q$ rounds of exploration is given by
\begin{align}
\mathcal R_S = \widehat{\L}_S( \textrm{max}\{\widehat{q}^*_S(q), q\}; q).\label{reg}
\end{align}
Our adaptive explore-then-commit algorithm with an optimally constructed LRMC estimator is presented in \Cref{alg2}. The algorithm starts (line 1) by collecting an initial set of model samples to estimate the model costs and the maximum number of exploration samples (line 2). The while loop then implements the main components of the exploration phase. Specifically, lines 4--6 compute the current estimates of the components of the MSE in \eqref{eq:estimated-loss} for every subset of models using the current exploration samples. Similar to \cite{xu2022bandit}, we use a regularization term, $\alpha_q\to 0$ as $q\to\infty$, which encourages early exploration and is asymptotically negligible. This term is often chosen to decay quickly (e.g., exponentially in $q$) in practice. Line 8 finds the model subset with the smallest MSE. Lines 9--25 determine if more exploration is needed and increment the number of exploration samples accordingly. Since finding the optimal $S$ can be relatively expensive in exploration (as it requires checking for all $S\subseteq [n]$), we employ a bisection trick introduced in \cite{xu2021budget, han2023approximate} to accelerate computation. This trick reduces the number of outer loops by allowing $q$ to grow nonlinearly, but the exponential dependence on $n$ in each inner loop remains. One way to alleviate this complexity is to restrict attention to a subset of models rather than enumerating all subsets of $[n]$ to accelerate computation (see \Cref{sec:ice-sheet}). Since the costs are random, it is theoretically possible that the requested exploration samples cannot be drawn using the remaining budget. In this case, the algorithm is terminated. Line 26 updates the estimated cost of a single exploration based on the augmented exploration samples drawn in the previous steps. Finally, when the exploration phase ends, line 28 computes the high-fidelity mean using the improved LRMC estimator $\lrmc$ based on the selected subset $\widehat{S}^*$ of low-fidelity models using the remaining budget. 

For fixed $B$, the total exploration cost is $q\widehat{c}_\ex \leq B$. Since $\widehat{c}_\ex \to c_\ex>0$ a.s. by the law of large numbers, $q$ is bounded from above a.s. Therefore, \Cref{alg2} terminates in finitely many steps a.s. Although \Cref{alg2} allows for random costs, we do not account for this extra randomness when computing the conditional MSE. The reason for this simplification is that otherwise the costs would appear in the denominator, making an exact derivation impossible. Instead, we use their expected values in the analysis. From this perspective, the analysis of random costs is similar to that of deterministic costs, except for the possibility of early termination in lines 14 and 21. Under mild assumptions, such terminations do not occur for all sufficiently large $B$ a.s.

\begin{algorithm}
\scriptsize
\hspace*{\algorithmicindent} \textbf{Input}: $B$: total budget, \textsf{Exploration}: True, $\alpha_t\downarrow 0$: regularization parameters\\
    \hspace*{\algorithmicindent} \textbf{Output}: an estimator for $\mu_0$
 \begin{algorithmic}[1]
 \STATE collect $q = n+2$ independent samples of $(Q_0, \ldots, Q_n)$ for exploration
\STATE compute $\widehat{c}_{\ex} =\widehat{c}_{\afs}$ and the maximum exploration round $M = \lfloor B/\widehat{c}_{\ex}\rfloor$
 \WHILE{\textsf{Exploration} $=$ True}
 \FOR{$S\subseteq [n]$}{
  \STATE{compute $\widehat{k}_q(S)$ and $\widehat{\gamma}_q(S)$ and set $\widehat{k}_q(S) \gets \widehat{k}_q(S) +\alpha_q$ 
  }
 \STATE{compute the optimal loss $\mathcal R_S$ using \eqref{haozi}-\eqref{reg}}
}
 \ENDFOR
 \STATE {compute the optimal model $\widehat{S}^*(q) = \argmin_{S\subseteq [n]}\mathcal R_S$}
\IF{$\widehat{q}^*_{\widehat{S}^*(q)}(q)>2q$}
  \STATE \textbf{try} to collect $q$ additional exploration samples
  \IF{successful}
    \STATE $q \gets 2q$
  \ELSE
    \STATE \textbf{break}
  \ENDIF
\ELSIF{$q<\widehat{q}^*_{\widehat{S}^*(q)}(q)\leq 2q$}
  \STATE \textbf{try} to collect $(\lceil\frac{q + \widehat{q}^*_{\widehat{S}^*(q)}(q)}{2}\rceil - q)$ additional exploration samples
  \IF{successful}
    \STATE $q \gets \lceil\frac{q + \widehat{q}^*_{\widehat{S}^*(q)}(q)}{2}\rceil$
  \ELSE
    \STATE \textbf{break}
  \ENDIF
\ELSE
  \STATE $\textsf{Exploration} \gets$ False
\ENDIF
 \STATE compute $\widehat{c}_{\ex}$ and update the maximum exploration round counter $M = \lfloor B/\widehat{c}_{\ex}\rfloor$
 \ENDWHILE
 \STATE {return the LRMC estimator $\lrmc = \widehat{y}_{\widehat{S}^*(q)}^\top\bt_{\widehat{S}^*(q)}$ associated with $\widehat{y}_{\widehat{S}^*(q)}=(1, \mu^\top_{\widehat{S}^*(q)})^\top$} \label{alg:estimator}
 \end{algorithmic}
\caption{AETC-OPT algorithm for multi-fidelity approximation} 
\label{alg2}
\end{algorithm}

Technically, computing MLBLUEs (either for $\widehat{\gamma}_q(S)$ or $\lrmc$) requires the covariance matrix across low-fidelity models. However, one can use its empirical estimator from the exploration phase as a substitute, which does not incur additional cost. This approximation introduces additional variance that has an asymptotically negligible effect on the final MSE. We will investigate this theoretically and numerically in \Cref{sec:exploitationBLUE} and \Cref{sec:numerical}, respectively.

\section{Theoretical Analysis}\label{sec:exploitationBLUE}

In this section, we analyze the AETC-OPT algorithm as described in \Cref{alg2}. First, we verify that choosing $\widehat{\mu}_S$ as an MLBLUE for $\mu_S$ in \Cref{sec:algorithm} fits into the proposed bandit-learning framework in \Cref{sec:new_MSE}. Second, we prove asymptotic results for \Cref{alg2} under appropriate assumptions. Third, we demonstrate the robustness and optimality of $\lrmc$ used in the exploitation phase of \Cref{alg2} by connecting it to a broad class of multi-fidelity estimators based on ACVs~\cite{Gorodetsky_2020}, as well as to the MLBLUE for the high-fidelity mean under the optimal sample allocation~\cite{Schaden_2020}. 

\subsection{Properties of $\widehat{\mu}_S$}\label{sec:4.1}
Recall that $\widehat{\mu}_S$ is the MLBLUE in \eqref{eq:BLUE_estimator} with $\{m_T\}_{T\subseteq S}$ selected as the optimal allocation determined using the estimated sketch $\widehat{b}_S$. Here, we allow $\widehat{\mu}_S$ to be computed using either the oracle covariance matrix $\Sigma_S$ or its empirical estimator $\widehat{\Sigma}_S$ from the exploration phase.
We begin by verifying the following properties of $\widehat{\mu}_S$:
\begin{itemize}
\item \Cref{4210}: $\widehat{\mu}_S$ is exploration-unbiased.
\item \Cref{4211}: $\widehat{\mu}_S$ satisfies the asymptotic scaling property \eqref{gamma} with some finite $\gamma_S$.
\item \Cref{4212}: $\gamma_S$ is consistently estimable using the exploration data.
\end{itemize}
The first two properties are required to derive the asymptotic MSE in \eqref{asympform} for $\lrmc$ in the exploitation phase, as shown in \Cref{lemma:bl-asymptotic-mse}. The last property is needed for the implementation of AETC-OPT and proof of the consistency and optimality of \Cref{alg2}  (\Cref{thm:AETC}). Detailed proofs of these technical results are deferred to \Cref{sec:supp4.1}. To proceed, we need some tail-decay assumptions on the random variables, which may be stronger than necessary but are adopted for convenience.  
\begin{Def} 
For $\alpha\geq 1$, the $\alpha$-Orlicz norm of a random variable $Z$ is defined as
\begin{align}
\|Z\|_{\psi_\alpha} = \inf\left\{C>0: \E\left[\exp\left(\frac{|Z|^\alpha}{C^\alpha}\right)\right]\leq 2\right\}.
\end{align}
In particular, $Z$ is called \emph{sub-Gaussian} if $\|Z\|_{\psi_2}<\infty$ \cite[Chapter 2]{Vershynin2018}.
\end{Def}

In the rest of the section, we verify the above properties step by step under the following technical assumptions:

\begin{Ass}\label{subgaussian}
$\max_{S\subseteq [n]}\|\e_S\|_{\psi_2}<\infty$.   
\end{Ass}

\begin{Ass}\label{subexp}
$\max_{i\in [n]}\E[|Q_i|^3]<\infty$.  
\end{Ass}

\begin{Ass}\label{myass}
For every $S\subseteq [n]$, the optimal objective in the relaxation of \eqref{optint} with $\afs$ replaced by $S$ (i.e., $\mathcal M = \{m_T\}_{T\subseteq S}\in\R_+^{2^{s}-1}$), as a function of $(a, \Sigma, c)$, is continuous at $(b_S, \Sigma_S, c_S)$. 
\end{Ass}

The exploration unbiasedness of $\widehat{\mu}_S$ is given by the following lemma. 

\begin{Lemma}[Exploration unbiasedness of $\widehat{\mu}_S$]\label{4210}
$\E[\widehat{\mu}_{S}\mid \widehat{b}_S,Z_\lfs] = \mu_S$. 
\end{Lemma} 

Under \Cref{subgaussian}–\ref{myass}, we prove that $\widehat{\mu}_S$ satisfies the asymptotic scaling property~\eqref{gamma} by explicitly identifying the $\gamma_S < \infty$.

\begin{Lemma}[Existence of $\gamma_S \in (0, \infty)$]\label{4211}
Under the linear model assumption~\eqref{lr} and \Cref{subgaussian}-\ref{myass}, for every $S\subseteq [n]$, 
\begin{align}
& \lim_{\min\{q, B_\ext\}\to\infty}B_\ext\cdot \E\left[\widehat{b}_S^\top\Cov[\widehat{\mu}_{S}\mid \widehat{b}_S,Z_\lfs]\widehat{b}_S\mid Z_\lfs\right]\nonumber\\
 =&\  \min_{\substack{\mathcal M = \{m_T\}_{T\subseteq S}\in\R_+^{2^s-1}\\ \sum_{T\subseteq S}c_T m_T\leq 1}}b_S^\top \left(\sum_{T\subseteq S}m_T R_T^\top \Sigma_T^{-1}R_T\right)^{-1}b_S&a.s.,\label{des}
\end{align}
where $B_\ext = B - c_\ex q$. 
\end{Lemma}

\Cref{4211} identifies an explicit form of $\gamma(S)$ as 
\begin{align}
\gamma(S) = \min_{\substack{\mathcal M = \{m_T\}_{T\subseteq S}\in\R_+^{2^s-1}\\ \sum_{T\subseteq S}c_T m_T\leq 1}}b_S^\top \left(\sum_{T\subseteq S}m_T R_T^\top \Sigma_T^{-1}R_T\right)^{-1}b_S.\label{gamama}
\end{align}  
A straightforward estimator for $\gamma_S$ using estimated information is the following:
\begin{align}
\widehat{\gamma}_q(S)  = \min_{\substack{\mathcal M = \{m_T\}_{T\subseteq S}\in\R_+^{2^s-1}\\ \sum_{T\subseteq S}\widehat{c}_T m_T\leq 1}}\widehat{b}_S^\top \left(\sum_{T\subseteq S}m_T R_T^\top \widehat{\Sigma}_T^{-1}R_T\right)^{-1}\widehat{b}_S,\label{et}
\end{align} 
where $\widehat{\Sigma}_T$ and $\widehat{c}_T$ are the empirical estimators for $\Sigma_T$ and $c_T$ using exploration data defined in \eqref{haozi}. The next lemma shows that $\widehat{\gamma}_q(S)$ is a consistent estimator for $\gamma(S)$ as $q\to\infty$. 

\begin{Lemma}[Estimability of $\gamma_S$]\label{4212}
Under Assumptions \ref{subgaussian}-\ref{myass}, $\lim_{q\to\infty}\widehat{\gamma}_q(S) = \gamma(S)$ a.s. 
\end{Lemma}

\subsection{Algorithmic Analysis}\label{sec:alga}

Under the same choice of $\widehat{\mu}_S$ in \Cref{sec:4.1}, the following result can be proved for \Cref{alg2}:

\begin{Th}\label{thm:AETC}
Let $q(B)$ be the number of exploration rounds chosen by \Cref{alg2} under budget $B$, and $S(B)$ be the model for exploitation. Under the linear model assumption~\eqref{lr} and Assumptions \ref{subgaussian}-\ref{myass}, if $\lim_{q\to\infty}\alpha_q = 0$ and $\{\C_i\}_{i\in\afs}$ are positive and uniformly bounded above, then a.s., 
\begin{subequations}
\begin{align}\label{rr1}
  \lim_{B\to\infty}\frac{q(B)}{q^*_{S^*}} &= 1, \\
  \lim_{B\to\infty}S(B) &= S^*,\label{rr2}
\end{align}
\end{subequations}
where $S^*$ and $q^*_{S^*}$ are the model choice and exploration round count given oracle information defined in \eqref{pj2} and \eqref{pj1}, respectively.
\end{Th}
 The consistency of the output of \Cref{alg2} follows as a corollary of \Cref{thm:AETC}:

\begin{Cor}\label{cvgh}
The LRMC estimator $\lrmc$ produced by \Cref{alg2} converges to $\mu_0$ as $B\to\infty$ a.s. 
\end{Cor}
The proofs of \Cref{thm:AETC} and \Cref{cvgh} are modified based on those in \cite{xu2022bandit, xu2021budget, han2023approximate} and deferred to \Cref{sec:supp4.2}.

\subsection{LRMC, ACVs, and MLBLUEs}\label{sec:lam}

This section presents a quantitative analysis to establish firm connections between LRMC, ACVs, and MLBLUEs. In particular, the LRMC estimator $\lrmc$ is used in the exploitation phase of AETC-OPT to produce a final estimate for the high-fidelity mean. We provide evidence that the MSE of $\lrmc$ produced by AETC-OPT is commensurate with that of the optimal MLBLUE in \cite{Schaden_2020} computed with oracle statistics, which is empirically observed in \Cref{sec:numerical}. Proofs of all technical results in this section are deferred to \Cref{sec:supp4.3}. 

\begin{figure}[htbp]
\begin{center}
\includegraphics[width=0.75\textwidth]{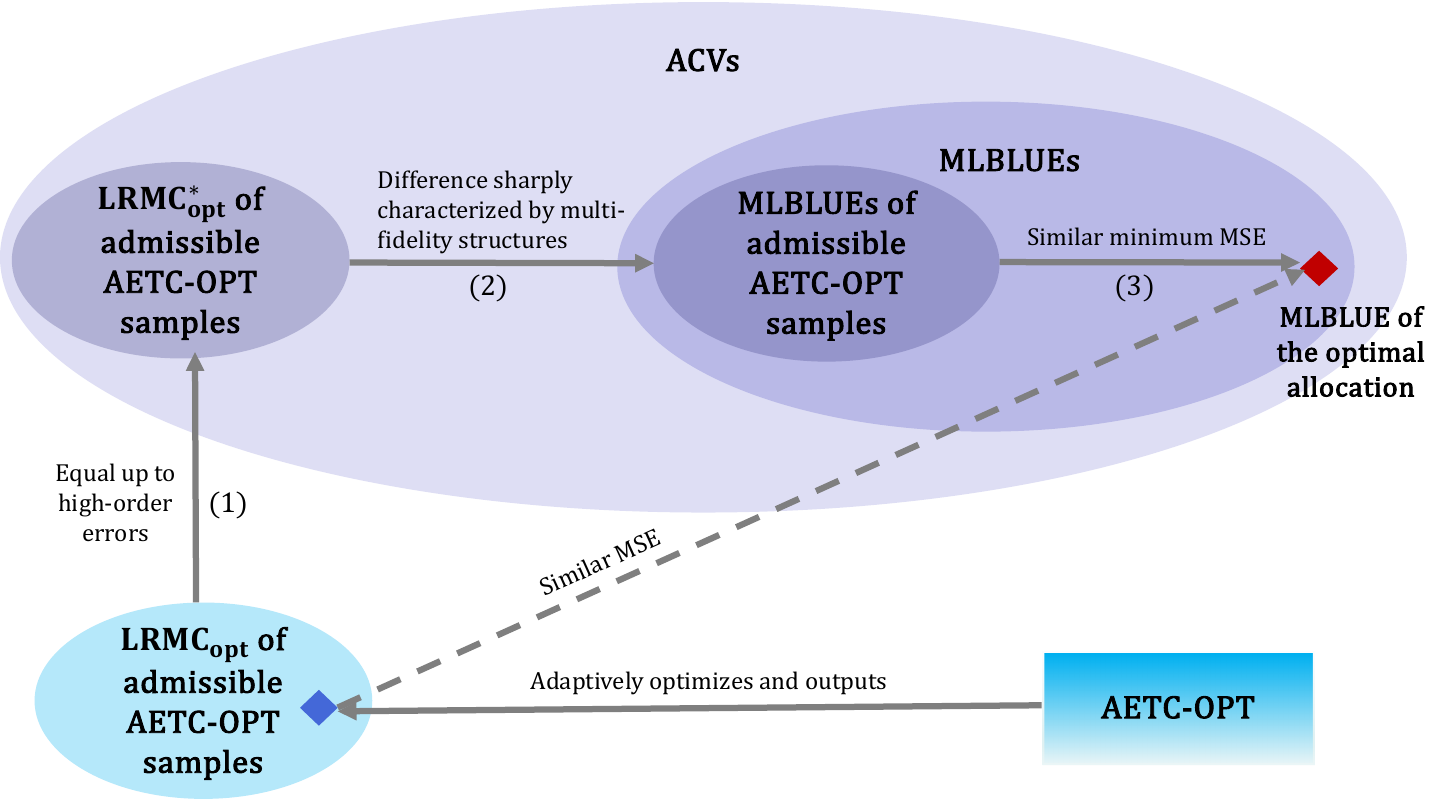}
  \caption{\small Demonstration of the relationship between $\lrmc$, $\lrmcstar$, ACVs, and MLBLUEs. In particular, the AETC-OPT algorithm adaptively optimizes and outputs an $\lrmc$ (the blue diamond) for the high-fidelity mean. Up to a first-order error whose coefficient is sharply characterized by multi-fidelity structures, the $\lrmc$ coincides with the MLBLUE under the same exploration and exploitation samples. Moreover, this MLBLUE exhibits a comparable MSE to the MLBLUE for the high-fidelity mean under the globally optimal sample allocation (the red diamond). Specifically, the inclusion of $\lrmcstar$ within ACVs is due to \Cref{lemma:acv}; Arrow (1) is due to \Cref{addback}, Arrow (2) is due to \Cref{khty}, and Arrow (3) is due to \Cref{ajan}. The dashed arrow is empirical but supported by the evidence provided in (1)--(3). }\label{fig:comparison}
  \end{center}
\end{figure}

\Cref{fig:comparison} offers a visual summary of our results. (1) \Cref{sec:lrmc_vs_acv} establishes that an idealized version of $\lrmc$, denoted by $\lrmcstar$, is a type of ACVs for the high-fidelity mean. Meanwhile, $\lrmc$ and $\lrmcstar$ differ by a higher-order term. These results justify the robustness of $\lrmc$ and suggest studying $\lrmcstar$ in lieu of $\lrmc$ to better understand the properties of the latter. (2) \Cref{sec:lrmc_vs_blue} studies the relationship between $\lrmcstar$ in the exploitation phase of AETC-OPT and the MLBLUE under the same allocation samples. We show that their difference is sharply characterized by multi-fidelity structures. (3) Moreover, under suitable cost assumptions, the MSE of the MLBLUE for the high-fidelity mean with the optimal allocation among admissible sample allocations under the AETC-OPT algorithm is comparable to that of the MLBLUE with the globally optimal allocation in \cite{Schaden_2020}. Collectively, these new results shed light on how AETC-OPT augments the original AETC procedure by improving the final estimation toward optimality. 

\subsubsection{LRMC vs ACVs}\label{sec:lrmc_vs_acv}

In this section, we compare the form of the ACV estimator with that of $\lrmc$ used in \Cref{alg2}. For completeness, we first give a brief introduction to ACV estimators. 

Control variate methods are a principled approach to variance reduction \cite{glasserman2004monte, owen2013monte}. Given $\mu_\lfs = \E[Q_\lfs]$ is known, the control variate method considers the following type of estimators for $\mu_0=\E[Q_0]$:
\begin{align}
    \widehat{\mu}_{0,\textrm{cv}}(\alpha) &= \frac{1}{m_0}\sum_{\ell\in M_0}W_{0,\ell} -\sum_{i\in \lfs}\alpha_i\left(\frac{1}{m_i}\sum_{\ell\in M_i}W_{i,\ell}-\mu_i\right) \quad\quad\quad \alpha\in\R^{n},\label{lbhy} \\
    &= \Bar{\mu}_0 (M_0) -\sum_{i\in \lfs}\alpha_i\left(\Bar{\mu}_i (M_i)-\mu_i\right) \nonumber \\
    &= \Bar{\mu}_0 (M_0) - \alpha^\top \left(\Bar{\mu}_\lfs (M_\lfs)-\mu_\lfs\right)\nonumber
\end{align}
where $W_{i,\ell}$ is the $\ell$th sample of $Q_i$, the $i$th model fidelity, and $M_0$ and $\{M_i\},i\in \lfs$ are sets of sample indices for the high-fidelity model and low-fidelity models, respectively. Let $m_0 = |M_0|$ and $\{m_i=|M_i|\},i\in \lfs$ be the sizes of the sample sets.

However, the means of the low-fidelity models, $\mu_\lfs$, are not known in practice. The ACV framework proposes estimating $\mu_\lfs$ \cite{Gorodetsky_2020} with additional sets of sample indices $\{P_i\},i\in \lfs$, 
\begin{align}
    \widehat{\mu}_{0,\mathrm{acv}}(\alpha) &= \Bar{\mu}_0 (M_0) - \alpha^\top \left(\Bar{\mu}_\lfs (M_\lfs)-\widehat{\mu}_\lfs (P_\lfs) \right) \label{eq:ACVdefinition}.
\end{align}
The sample index sets $M_0$, $M_\lfs = \{M_i\},i\in \lfs$, and $P_\lfs = \{P_i\},i\in \lfs$ are called the sample allocation, $\mathcal M$, and must satisify
\begin{align}
    m_0 c_0 + \sum_{i\in \lfs} | M_i \cup P_i | c_i \leq B.
\end{align}

ACV estimators are unbiased, so their MSE is equivalent to variance. The weights $\alpha$ can be chosen to minimize the variance of the estimator
\begin{align*}
    \alpha^* &= \Cov\Bigl[\Bar{\mu}_0 (M_0),~ \Bar{\mu}_\lfs (M_\lfs) -\widehat{\mu}_\lfs (P_\lfs) \Bigr]\Cov\Bigl[\Bar{\mu}_\lfs (M_\lfs) -\widehat{\mu}_\lfs (P_\lfs)\Bigr]^{-1}\\     &= \left[F^\top \circ \Cov[Q_0,Q_\lfs]\right]\left[G \circ \Cov[Q_\lfs]\right]^{-1}
\end{align*}
where $\circ$ denotes the element-wise Hadamard product, and $F\in\R^{n}$ and $G\in\R^{n\times n}$ are deterministic matrices that can be found based on the sample allocation $\mathcal M$ \cite{Bomarito_2022}.

The following lemma shows that the LRMC estimator $\lrmc$ used in \Cref{alg2} can be identified as a class of ACV estimators in \eqref{eq:ACVdefinition} with $\alpha$ estimated by exploration samples. 

\begin{Th}\label{lemma:acv}
Recall $\lrmc = \widehat{y}_S^\top\widehat{\beta}_S$ in \Cref{alg2}, where $\widehat{y}_S = (1, \widehat{\mu}^\top_S)^\top$ and $\widehat{\beta}_S = (\widehat{a}_S, \widehat{b}^\top_S)^\top$ is defined in \eqref{lse}, with $\widehat{a}_S, \widehat{b}_S$ denoting the estimated intercept and low-fidelity coefficients, respectively. Then $\lrmc$ can be identified as a type of ACV estimator \eqref{eq:ACVdefinition} with the coefficients vector $\alpha$ replaced by an exploration-dependent estimator. 
\end{Th}

\Cref{lemma:acv} shows that $\lrmc$ can be approximately identified as an ACV estimator. Since ACVs operate under less restrictive assumptions than joint linear regression, this highlights the practical robustness of $\lrmc$.

Due to the dependence between $\widehat{b}_S$ and $\Bar{\mu}_0$ (since both use exploration samples), $\lrmc$ is no longer unbiased for $\mu_0$. Nonetheless, the additional MSE is asymptotically negligible assuming $\widehat{\mu}_S$ is at least as accurate as $\Bar{\mu}_S$. Establishing this quantitative result requires the linear model assumption~\eqref{lr}. 

\begin{Th}\label{addback}
Consider the true ACV obtained by replacing $\widehat{b}_S$ by its true value:
\begin{align}
\lrmcstar=\Bar{\mu}_0- b_S^\top(\Bar{\mu}_S- \widehat{\mu}_S).\label{hute}
\end{align}
If the linear model assumption \eqref{lr} and \Cref{subexp} hold and $\Cov[\widehat{\mu}_S\mid\widehat{b}_S, Z_\lfs]\preceq \Cov[\Bar{\mu}_S]=\Sigma_S/q$, then $\E[(\lrmcstar-\lrmc)^2\mid Z_\lfs] = o(\V[\lrmcstar])$ a.s. 
\end{Th}

\subsubsection{LRMC vs MLBLUEs}\label{sec:lrmc_vs_blue}

Thanks to \Cref{addback}, we can now study $\lrmcstar$ in place of $\lrmc$ in the exploitation phase of the AETC-OPT algorithm to obtain cleaner results. Fixing the exploration samples, exploitation index set $S$, and exploitation samples over $S$ (joint samples of $Q_T$ for $T\subseteq S$), the idealization $\lrmcstar$ of $\lrmc$ and the MLBLUE for $\mu_0$ with respect to both the exploration and exploitation samples match up to error terms sharply characterized by multi-fidelity structures.  

\begin{Th}\label{khty}
Fix $q$ exploration samples $\{Q_{\afs, i}\}_{i\in [q]}$, the index set $S\subseteq [n]$ with $|S|=s$, and exploitation samples $\{W_{T, \ell}\}_{T\subseteq S, \ell\in [m_T]}$. Let $\widehat{\mu}_\blue$ be the corresponding MLBLUE for $\mu_\afs$ defined in \eqref{eq:BLUE_estimator}. Define 
\begin{align}
&\Delta = q\Xi^{-1}\Sigma_S^{-1} &\Xi = \sum_{T\subseteq S}m_TR_T^\top\Sigma_T^{-1}R_T, \label{shaoxin}
\end{align} 
where $R_T\in\{0, 1\}^{|T|\times s}$ is the restriction matrix to the index set $T\subseteq S$, and consider the first component of $\widehat{\mu}_{\blue}$, denoted by $\widehat{\mu}_{0,\blue}$. If $\Delta$ is small, i.e., $\|\Delta\|_2<1/2$, then
\begin{align}
\widehat{\mu}_{0,\blue} = \lrmcstar + \mathcal O(\|\Delta\|_2\|\Bar{\mu}_S - \widehat{\mu}_S\|_2),\label{kshsy}
\end{align}
where $\lrmcstar$ is defined in \eqref{hute}. 
\end{Th}

\begin{Rem}
To understand the magnitude of $\|\Delta\|_2$, we consider a special case of uniform exploitation, i.e., $m_T>0$ only if $T=S$, which is a suboptimal allocation and serves as an upper bound.  In this case, $\Xi = m_S\Sigma^{-1}_S$, so $\|\Delta\|_2 = q/m_S$ represents the ratio between the size of exploration samples and exploitation samples, which is $\ll1$ in many multi-fidelity problems.
\end{Rem}

The MLBLUE to which $\lrmcstar$ is compared in \Cref{khty} is with respect to both allocation samples $\{Q_{\afs, i}\}_{i\in [q]}$ in exploration and $\{W_{T, \ell}\}_{T\subseteq S, \ell\in [m_T]}$ in exploitation. In the AETC-OPT algorithm, the allocation only allows the high-fidelity model to interact with all the low-fidelity models, but not with any proper subset of them. Theoretically, in addition to non-uniform exploitation, one could also consider non-uniform exploration where high-fidelity samples intersect differently with the low-fidelity models. However, as we will see, such flexibility offers little benefit when the high-fidelity model is significantly more expensive than the total cost of the low-fidelity models combined.

\begin{Th}\label{ajan}
Fix $S\subseteq [n]$ with $|S|=s$. Denote by $\mse^*_S(B)$ and $\mse_S(B)$ the MSE of the MLBLUE under the optimal sample allocation in $\mathcal{P}(S)\cup\{T\subseteq \afs: 0\in T\}$ and $\mathcal{P}(S)\cup\{\afs\}$ under the same budget $B>0$, respectively. Then,
\begin{align*}
\mse_S(B)\leq \frac{c_\ex}{c_0}\cdot\mse^*_S(B).
\end{align*}
\end{Th}

The MSE of the MLBLUE with respect to the globally optimal allocation in \eqref{optint} corresponds to $\mse^*_S(B)$ with $S = \lfs$. This observation, combined with Theorems~\ref{lemma:acv}-\ref{ajan} and the model-selection aspect of the AETC-OPT algorithm (\Cref{thm:AETC}), yields the diagram in \Cref{fig:comparison}. In particular, it implies the near-optimality of the LRMC estimator produced by the AETC-OPT algorithm compared to the MLBLUE under the globally optimal allocation computed with oracle statistics. 

\section{Numerical Experiments}\label{sec:numerical}
This section summarizes a numerical comparison of the properties and performance of the adaptively constructed multi-fidelity estimators given by the AETC-OPT algorithm with other multi-fidelity estimators when applied to two PDE models with uncertain parameters. Specifically, we investigate the performance of two versions of the AETC-OPT algorithm for estimating the mean of QoIs in an elasticity model and an ice-sheet model. The first AETC-OPT estimator uses oracle covariance information of the selected low-fidelity models for exploitation, while the second uses an empirical estimator derived from exploration data. For all AETC-type algorithms, we set the regularization parameters to $\alpha_q = 4^{-q}$ by default unless stated otherwise. We compare these two versions of the AETC-OPT algorithm with the original AETC algorithm, a single-fidelity MC estimator, and MFMC and optimal MLBLUE estimators that were both impractically, provided oracle information (i.e., involving the high-fidelity model). 
The following list presents the exact details of each estimator considered.

\begin{itemize}[leftmargin=2.66cm]
  \item[(MC)] Classical MC, where the entire budget is expended over the high-fidelity model.
  \item[(MFMC)] MFMC \cite{Peherstorfer_2016}, which is provided with oracle covariance information of both the high and low-fidelity models. Given oracle estimation, we optimize the sample allocation numerically to avoid the modeling constraints required to minimize the sample allocation analytically.
   \item[(MLBLUE)] MLBLUE with the optimal sample allocation \cite{Schaden_2020}, which is computed using oracle covariance information of both the high and low-fidelity models. We optimize the sample allocation using semi-positive definite programming~\cite{Croci_WW_CMAME_2023}.
  \item[(AETC)] The LRMC estimator produced by AETC \cite{xu2022bandit}. 
  \item[(AETC-OPT)] The LRMC estimator produced by \Cref{alg2}.
  \item[(AETC-OPT-E)] The LRMC estimator produced by \Cref{alg2} with computations based on the estimated covariance information of the low-fidelity models from exploration data.
\end{itemize}
For estimators that have exploration stages that adaptively determine the number of pilot samples, one could reuse these pilot samples in exploitation at the potential cost of introducing additional bias in the procedure. In practice, we observe a negligible empirical impact on bias with this approach. Meanwhile, since our pilot sample sizes are relatively small compared to the samples in the exploitation phase, pilot sample re-use also provides negligible accuracy benefits in our examples. This phenomenon in the AETC setting has been observed in \cite[SM1]{xu2022bandit}. Because of this, and to be consistent with the explicit algorithm we have introduced, we present results \textit{without} re-use of pilot samples.

\subsection{Linear Elastic Displacement}\label{sec:elastic-pde}

The first part of our study focuses on estimating uncertainty in a parametric elliptic equation that describes the displacement of an elastic material under forcing $f(\bm{x})$~\cite{xu2022bandit}. This is achieved using a multi-fidelity ensemble of models, each employing different numerical discretizations to solve the equation. The uncertainty is introduced through a 4-term Karhunen–Loève expansion of a random field representing the elastic modulus. The expansion's coefficients are independent and identically distributed standard normal random variables.
We construct a multi-fidelity hierarchical model ensemble by coarsening the mesh parameter $h$ of a finite element solver with standard bilinear square isotropic finite elements defined on a rectangular mesh \cite{andreassen2011efficient}. The model with the finest mesh size is treated as the high-fidelity model $Q_0$, and the remaining four coarser-mesh models, $Q_1, \ldots, Q_4$, serve as the low-fidelity models. The scalar QoI is the structural \emph{compliance}, i.e., the energy norm of the displacement solution. The computational cost of each model is deterministic and inversely proportional to the mesh size squared ($h^2$); this corresponds to using a linear solver of optimal linear complexity. We normalize cost so that the lowest fidelity model has unit cost, i.e., $c_0 = 4096, c_1 = 64, c_2 = 16, c_3 = 4, c_4 = 1$. The correlations between the outputs of $Q_0$ and $Q_1, Q_2, Q_3, Q_4$ are $0.976$, $0.940$, $0.841$, $-0.146$, respectively. Additional details on the problem setup can be found in \Cref{sec:supp12}.  

\begin{figure}[htbp]
\begin{center}
\includegraphics[width=0.44\textwidth]{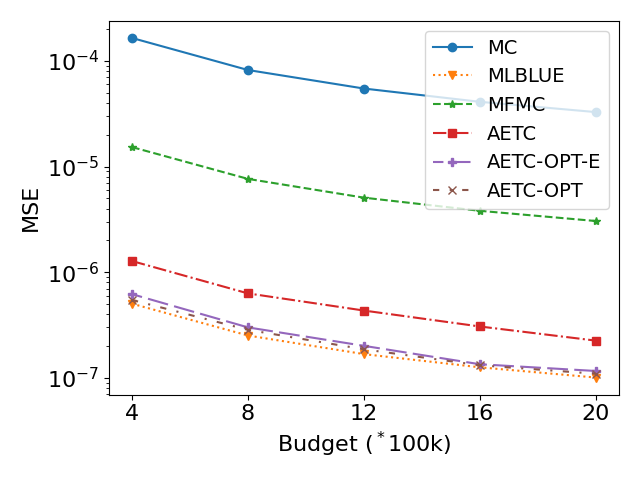}\hfill
\includegraphics[width=0.40\textwidth]{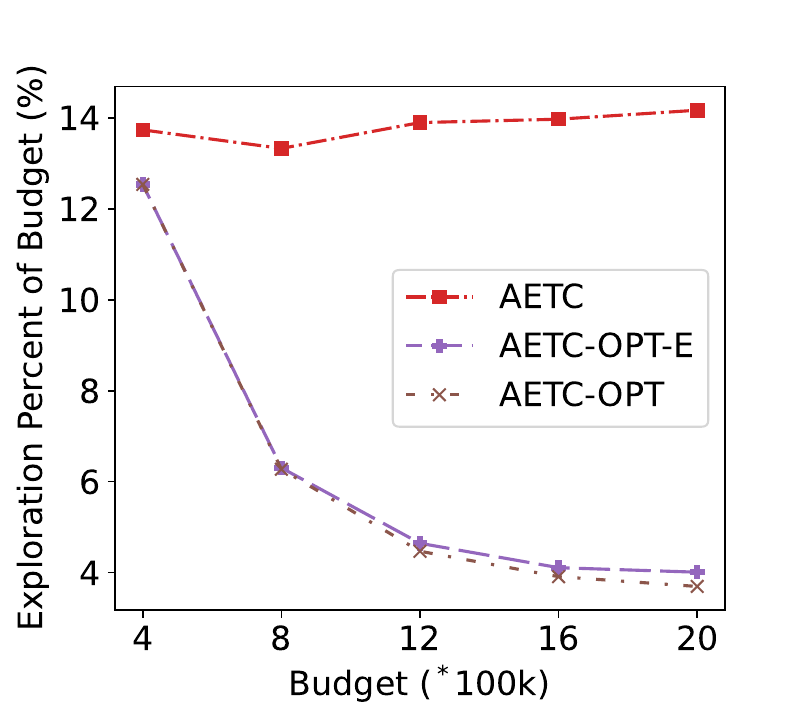}
\caption{\small ({\it Left}) Comparison of the empirical MSE of various estimators of compliance when selecting from 5 elasticity models as the total budget increases from $4\times 10^5$ to $2\times 10^6$. 
({\it Right}) The percentage of the total budget given to exploration sampling of AETC, AETC-OPT, and AETC-OPT-E across the budgets. }
  \label{fig:budgets}
  \end{center}
\end{figure}

The left plot of \Cref{fig:budgets} compares the MSE of 6 different methods for a range of total budgets $B$, specifically for equally spaced costs spanning $4\times 10^5$ to $2\times 10^6$, incremented by $4\times 10^5$ at a time. All multi-fidelity estimators were more accurate than the single-fidelity MC estimator, with MLBLUE having the smallest MSE. However, both MFMC and MLBLUE used oracle information and did not account for the cost of obtaining this oracle information. Consequently, MLBLUE represents a lower bound on the best possible performance of AETC-type methods. The AETC algorithm that used uniform exploitation did not have an MSE error close to this lower bound. Since AETC employs a strictly less flexible space of model allocations than MLBLUE, it is reasonable that its MSE is larger than that of MLBLUE. However, both AETC-OPT and AETC-OPT-E achieve MSE close to the lower bound while accounting for the cost of estimating the oracle statistics during the exploration phase. These results are consistent with our findings in \Cref{sec:exploitationBLUE}. 

The right plot of \Cref{fig:budgets} compares the percentage of the budget used for exploration sampling in the AETC-type algorithms. In this example, the AETC-OPT and AETC-OPT-E algorithms require fewer exploration samples than the AETC algorithm. This is because the optimal model selected by the AETC algorithm is smaller; the fewer exploration samples required by AETC-OPT suggest that the new loss function developed in this paper enables more nimble decision-making compared to AETC. The AETC-OPT/AETC-OPT-E algorithms selected the low-fidelity model set $S = \{1,2,3,4\}$ for exploitation, while the AETC algorithm chose $S = \{2,3,4\}$. Furthermore, there was little difference in the MSE error and the number of exploration/pilot samples used by the AETC-OPT and AETC-OPT-E algorithms, despite the former using oracle cross-model statistics of low-fidelity models.

\begin{figure}[htbp]
\begin{center}
\includegraphics[width=0.43\textwidth]{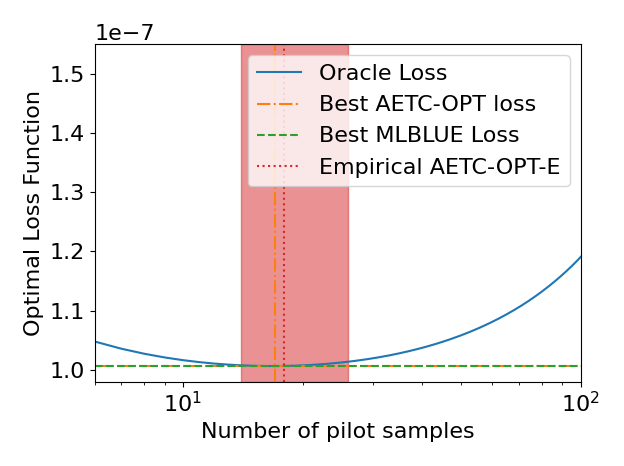}\hfill
\includegraphics[width=0.41\textwidth]{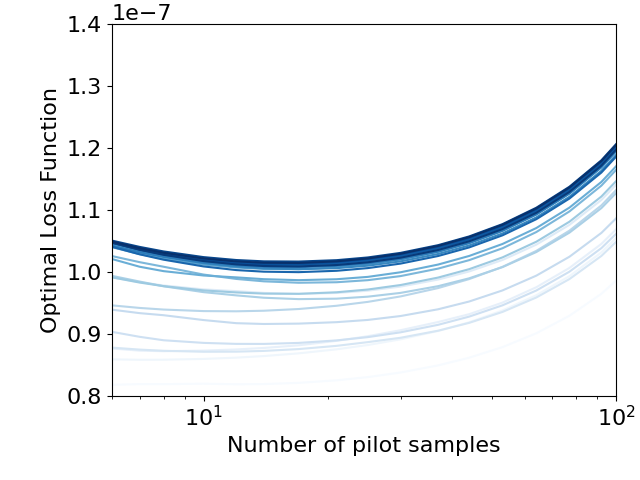}
\caption{\small ({\it Left}) The impact of the number of exploration samples on the MSE of the AETC-OPT algorithm in blue. The number of exploration samples chosen by AETC-OPT-E is denoted by the red-shaded region. ({\it Right}) Evolution of the estimated loss function as the AETC-OPT-E algorithm takes more exploration samples. Lighter curves represent loss functions with fewer exploration samples. }
  \label{fig:pilots}
  \end{center}
\end{figure}

The MSE of a multi-fidelity estimator is dependent on the number and cost of samples used to estimate the exploration statistics. Increasing the number of exploration/pilot samples used to calculate the estimated statistics increases the accuracy but decreases the budget that can be used for exploitation. The loss function given in \eqref{thisn} is a convex function that quantifies this trade-off. For a total budget of 2,000,000, \Cref{fig:pilots} displays the loss function of the exploration-exploitation trade-off for the linear-elasticity problem. The best possible MSE of the optimal MLBLUE estimator is given in the dashed green line, and the solid blue curve of the figure depicts the optimal loss value across all subsets computed using \eqref{xiaodaihua} with oracle statistics. The loss obtained by the AETC-OPT-E algorithm is very close to the optimal value of MLBLUE despite not using oracle information. Moreover, the median number of explorations/pilot samples in the dotted red line selected by AETC-OPT-E over 1,000 different random trials is close to optimal, with the 0.05-0.95 quantiles in the shaded region. Specifically, the median number of exploration/pilot samples intersects close to the minimum of the loss curve computed using \eqref{xiaodaihua}. While there is variability in the number of exploration/pilot samples chosen between the 0.05 and 0.95 quantiles, the MSE varies little. 

The plot on the right in \Cref{fig:pilots} displays the empirically estimated loss functions using the covariance information from the exploration samples as the AETC-OPT-E algorithm iteratively adds more exploration samples. The lightly shaded blue curves represent the estimated loss function given fewer exploration samples, while the darker blue curves represent the loss function given more exploration samples. As more exploration samples are drawn, the loss function curve converges to the truth, as seen by the light-shaded curves converging to the darker region.

\begin{figure}[htbp]
\begin{center}
\includegraphics[width=0.42\textwidth]{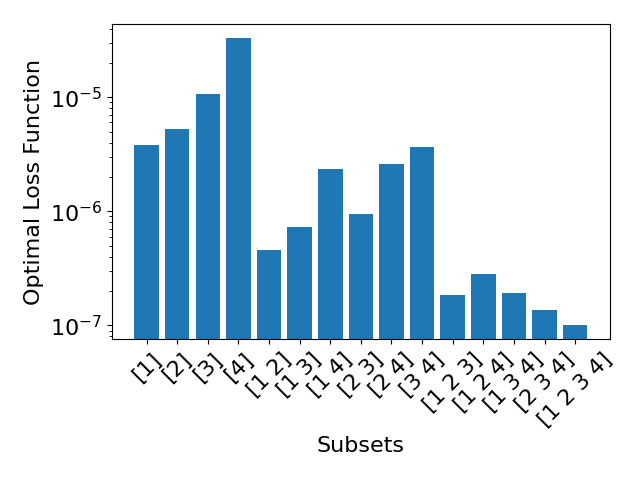}\hfill
\includegraphics[width=0.42\textwidth]{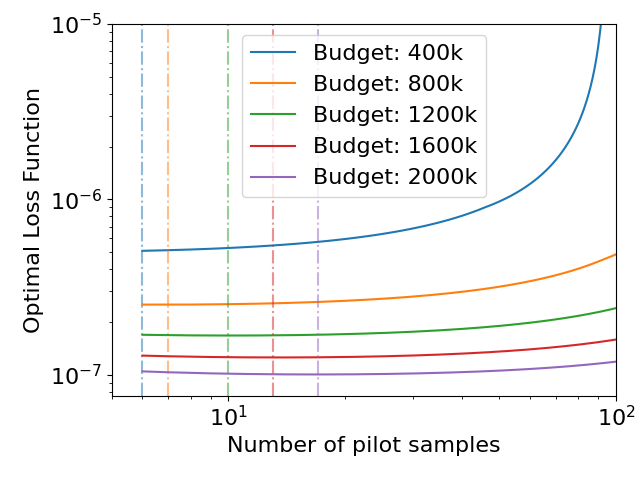}
\caption{\small ({\it Left}) The best oracle loss across all numbers of exploration samples of each subset of elasticity models. ({\it Right}) The loss as a function of the number of exploration samples and budgets is shown. 
 }
  \label{fig:subset}
  \end{center}
\end{figure}

The bar chart on the left in \Cref{fig:subset} displays the best loss value using the optimal number of exploration samples for each subset of low-fidelity models for the AETC-OPT algorithm. The smallest loss is achieved using $S=\{1,2,3,4\}$, all available low-fidelity models, which is what AETC-OPT and AETC-OPT-E selected every time in the 1,000 trials. This figure demonstrates the importance of performing optimal model selection, as the difference in MSE can be several orders of magnitude. The plot on the right in \Cref{fig:subset} plots the loss function for each budget. The vertical dot-dashed lines correspond to the optimal number of exploration samples for each budget. As the budget increases, the optimal number of exploration samples increases while the optimal loss value decreases. Not only does this figure demonstrate the decreasing MSE of AETC-OPT as a function of budget, but we also see that selecting the number of exploration samples is heavily dependent on budget. The AETC-OPT algorithm is able to adaptively identify the optimal number of exploration samples to minimize its MSE.

\begin{figure}[htbp]
\begin{center}
\includegraphics[width=0.43\textwidth]{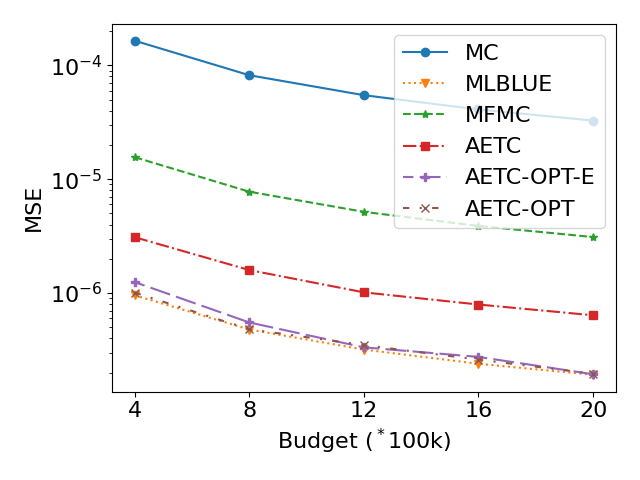}\hfill
\includegraphics[width=0.40\textwidth]{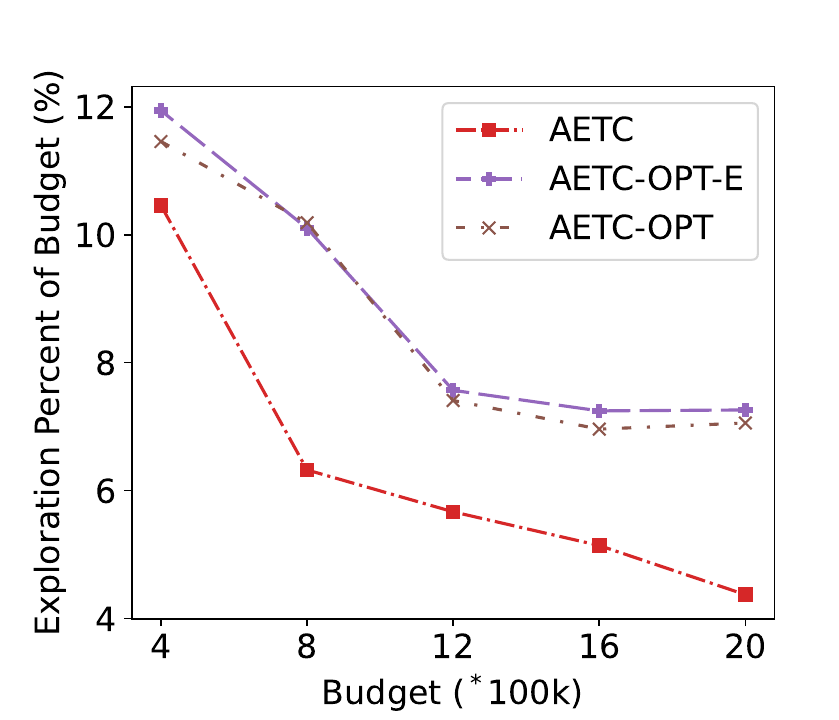}
\caption{\small 
({\it Left}) 
Comparison of the empirical MSE of various estimators of compliance when selecting from 4 elasticity models.
({\it Right}) The percentage of the total budget given to exploration sampling of AETC, AETC-OPT, and AETC-OPT-E across the budgets. The AETC-OPT algorithm significantly outperforms AETC due to larger exploitation costs. 
 }
  \label{fig:cut_test}
  \end{center}
\end{figure}

We now repeat the above experiment with the low-fidelity model $Q_2$ removed. The results are similar and reported in \Cref{fig:cut_test}. Without $Q_2$, AETC cannot select the original optimal model subset and thus has to spend more budget on exploitation to use the alternative surrogate $Q_1$. The new optimal model is $S = \{1,3,4\}$, which is the same as the one selected by AETC-OPT/AETC-OPT-E. The improvement of AETC-OPT/AETC-OPT-E over AETC is more stunning than before due to increased exploitation costs. Since AETC-OPT/AETC-OPT-E and AETC use the same model for exploitation, according to our analysis in \eqref{pj1}, the former will spend more budget on exploration (as they have a more efficient exploitation procedure reflected by a smaller $\gamma(S)$). This observation is verified in the last plot in \Cref{fig:cut_test}.

\subsection{Ice Sheet Mass Change}\label{sec:ice-sheet}

Quantifying uncertainty in ice sheets is critical for guiding future policy since the melting of land-based ice sheets is predicted to contribute significantly to future rises in sea level. The amount of sea-level rise, however, is subject to large amounts of uncertainty, necessitating efficient uncertainty quantification. Recent work~\cite{jakeman_PSHHHP_Egusphere_2024} showed that multi-fidelity estimators can effectively leverage the high correlation between each model’s prediction of mass change of the Humboldt Glacier in Greenland modeled between 2007 and 2100 using a single future scenario. However, the same study also identified that when used to estimate the uncertainty in mass change, the performance of the estimators was highly sensitive to the number of exploration samples. A bootstrapping procedure was used to quantify this sensitivity for a fixed number of exploration samples; however, the results suggested that an algorithm such as AETC-OPT-E is needed to objectively choose the number of exploration samples. Consequently, in this section, we document the performance of applying AETC-OPT/AETC-OPT-E to estimate the mean mass change from Humboldt Glacier.

\begin{figure}[htbp]
\begin{center}
\includegraphics[width=0.55\textwidth]{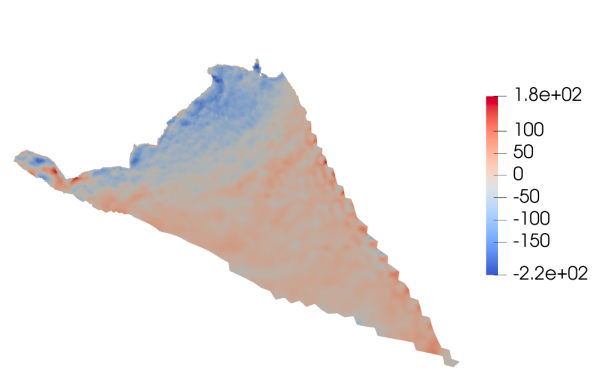}
\caption{\small The difference between the initial and final mass of Humboldt Glacier at year 2100 for a random sample from the posterior distribution of the basal friction field. The figure is reused from~\cite{jakeman_PSHHHP_Egusphere_2024} with the author's permission. 
 }
  \label{fig:ice-sheet}
  \end{center}
\end{figure}

Following~\cite{jakeman_PSHHHP_Egusphere_2024}, we investigate the use of 13 different models of varying computational cost and accuracy to compute ice-sheet mass change. The models were based on finite-element models of two differing physics models and varying mesh and time-step sizes.  Each model is parameterized by an 11,536-dimensional discretization of a Gaussian random field representing the basal friction between the land mass and ice sheet.  The prior distribution for the friction parameters was conditioned on observational data using Bayesian inference. Samples from the resulting posterior distribution are then used to predict the total mass change in year 2100. \Cref{fig:ice-sheet} depicts the computational domain and the change in the mass at the final time computed using the highest fidelity model for a random sample from the posterior. 

Exactly, replicating the study in~\cite{jakeman_PSHHHP_Egusphere_2024}, the high-fidelity model was a mono-layer higher-order (MOLHO) model~\cite{dosSantos2022} and a Shallow-Shelf Approximation (SSA) model~\cite{morland_johnson_1980, weis_1999}. The MOLHO model incorporates simplifications based on the observation that ice sheets are typically shallow, meaning their horizontal extent is much greater than their thickness. On the other hand, the SSA model makes the additional assumption that the horizontal components of velocity do not vary with thickness and is typically considered less accurate than the MOLHO model. Both physics models were implemented using the same code base, which utilized a Python wrapper for the FEniCS finite-element library~\cite{fenics2015}.  The relative computational costs of the models are summarized in Table 2. For a complete description of the uncertainty quantification problem formulation used in this study, we refer the reader to~\cite{jakeman_PSHHHP_Egusphere_2024}.

\begin{figure}[!htb]
    \centering
    \resizebox{0.95\textwidth}{!}{
    \begin{minipage}[b]{.30\textwidth}
        \centering
        \captionof{table}{The computational costs of the ice-sheet models used in this study.}
        \resizebox{\textwidth}{!}{
        \begin{tabular}{ |c|c|c| } 
         \hline
         Index & Name & Cost (s)\\
         \hline
         0 & MOLHO$_{1,9}$ & 15489.2 \\ 
         1 & MOLHO$_{1,36}$ & 3727.4 \\ 
         2 & MOLHO$_{1.5,36}$ & 2248.23 \\
         3 & MOLHO$_{2,36}$ & 1489.3 \\
         4 & MOLHO$_{3,36}$ & 410.4 \\
         5 & SSA$_{1,36}$ & 1434.0 \\
         6 & SSA$_{1.5,36}$ & 850.9 \\
         7 & SSA$_{2,36}$ & 569.9 \\
         8 & SSA$_{3,36}$ & 161.5 \\
         9 & SSA$_{1,365}$ & 191.7 \\
         10 & SSA$_{1.5,365}$ & 110.7 \\
         11 & SSA$_{2,365}$ & 72.8 \\
         12 & SSA$_{3,365}$ & 20.2 \\
         \hline
        \end{tabular}
      }
        \label{fig:ice-costs}
        \vspace{1cm}
    \end{minipage}\hspace{0.00\textwidth}
    \begin{minipage}[b]{0.68\textwidth}
        \centering
        \includegraphics[trim={0 0 10pt 0},width=1\textwidth]{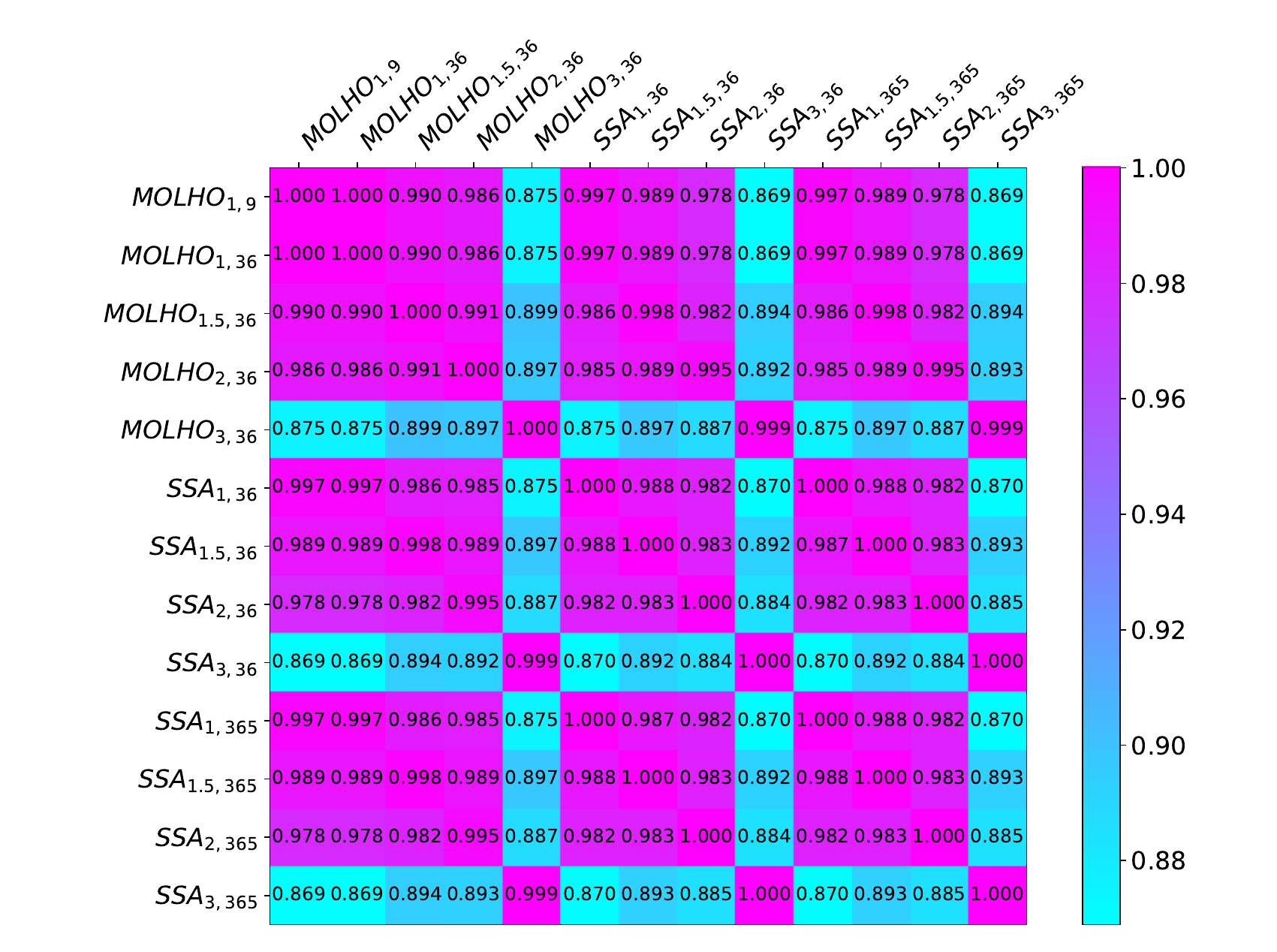}
        \caption{The correlations between the ice-sheet models used in this study.}
        \label{fig:ice-corr}
    \end{minipage}
    }
\end{figure}

Due to the expense of the ice-sheet model in this study, we generate 209 samples to perform a preliminary analysis of the predicted variance reduction of the AETC-OPT and AETC-OPT-E algorithms. Specifically, we run the exploration phases to find the predicted MSE and number of exploration samples required from \Cref{alg2}. However, the model's costs prevent us from executing the exploitation phase, but the estimates of the mean of the change in mass of Humboldt Glacier can be found in~\cite{jakeman_PSHHHP_Egusphere_2024}. The range of computational budget $B$ was chosen so that the AETC-OPT algorithms at the largest budget value used most of the available 209 samples. The median computational cost of each model is shown in \Cref{fig:ice-costs}, and the correlations between each model are shown in \Cref{fig:ice-corr}. Summing the computational costs \Cref{fig:ice-costs}, evaluating all models at one sample takes approximately five hours, and this total cost is dominated by the cost of simulating the high-fidelity model, which is approximately four hours. The subscripts in the model names correspond to the space and time discretizations of the solvers. The first subscript is the size of one spatial element in kilometers, and the second subscript is the length of the time step in days.

The study in~\cite{jakeman_PSHHHP_Egusphere_2024} found that 3 low-fidelity models were sufficient to substantially reduce the cost of estimating the mean mass change. Consequently, we set the maximum size of the low-fidelity model subsets considered by the AETC-OPT/AETC-OPT-E algorithms to four, resulting in  $\sum_{i=1}^4 {12 \choose i}=793$ possible model subsets $S$. To improve our ability to report the findings of our analysis, we further restrict the model subsets by choosing the 20 subsets with the lowest oracle MSE (estimated from all 209 samples). 

\begin{figure}[htbp]
\begin{center}
\includegraphics[width=0.4\textwidth]{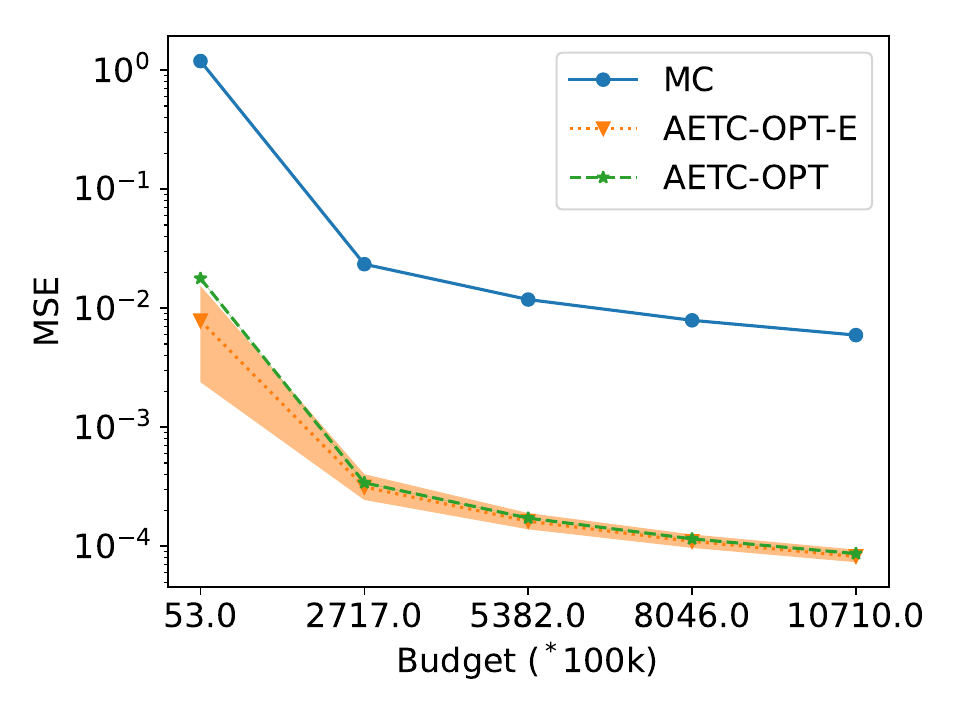}\hfill
\includegraphics[width=0.4\textwidth]{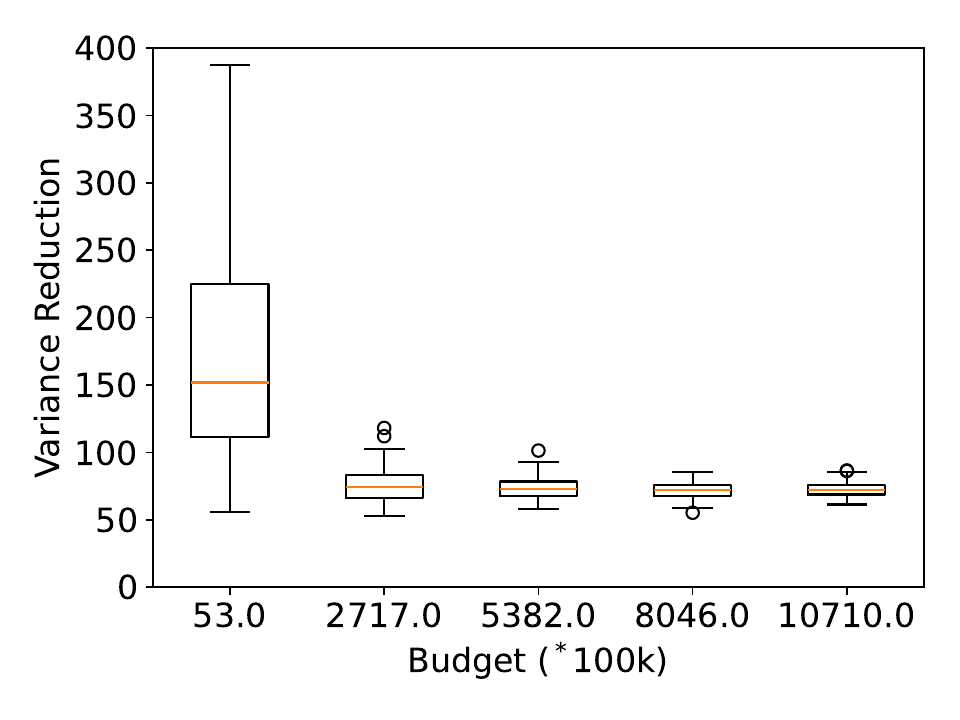}
\caption{\small {\it (Left)} Comparison of AETC-OPT-E with MC when used to estimate the mean mass loss from Humboldt-Glacier with 13 models. The dotted line represents the median loss, and the edges of the shaded region represent the 0.05-0.95 quantiles. The shaded region of AETC-OPT is too small to be seen. {\it (Right)} The variance reduction of using the AETC-OPT-E algorithm for increasing budgets. 
 }
  \label{fig:ice-budget}
  \end{center}
\end{figure}

The left panel of \Cref{fig:ice-budget} plots the optimal loss computed using \eqref{xiaodaihua} for 200 different estimators constructed by randomly bootstrapping with replacement the 209 samples available. The dotted orange line represents the median loss, and the edges of the shaded region represent the 0.05-0.95 quantiles. The AETC-OPT in green also contains a shaded region of quantiles, but it is too small to be plotted. Since AETC-OPT-E does not use all available samples to estimate the required statistics, it has a larger spread of MSE predictions. AETC-OPT is likely more accurate in its MSE prediction and has a much smaller spread while predicting a larger MSE than AETC-OPT-E. 

The associated variance reduction, defined as the variance of the equivalent-cost MC estimator divided by the variance of the LRMC estimator given by AETC-OPT-E, is shown on the right in \Cref{fig:ice-budget}. The plotted circles are outliers from the quantiles. The plot shows that as the budget increases, the variability in both the estimated loss and estimated variance reduction decreases. At the maximum budget, the predicted variance reduction is 72.4, which corresponds to a 72 times smaller variance compared to traditional MC estimation.

\begin{figure}[htbp]
\begin{center}
\includegraphics[width=0.42\textwidth]{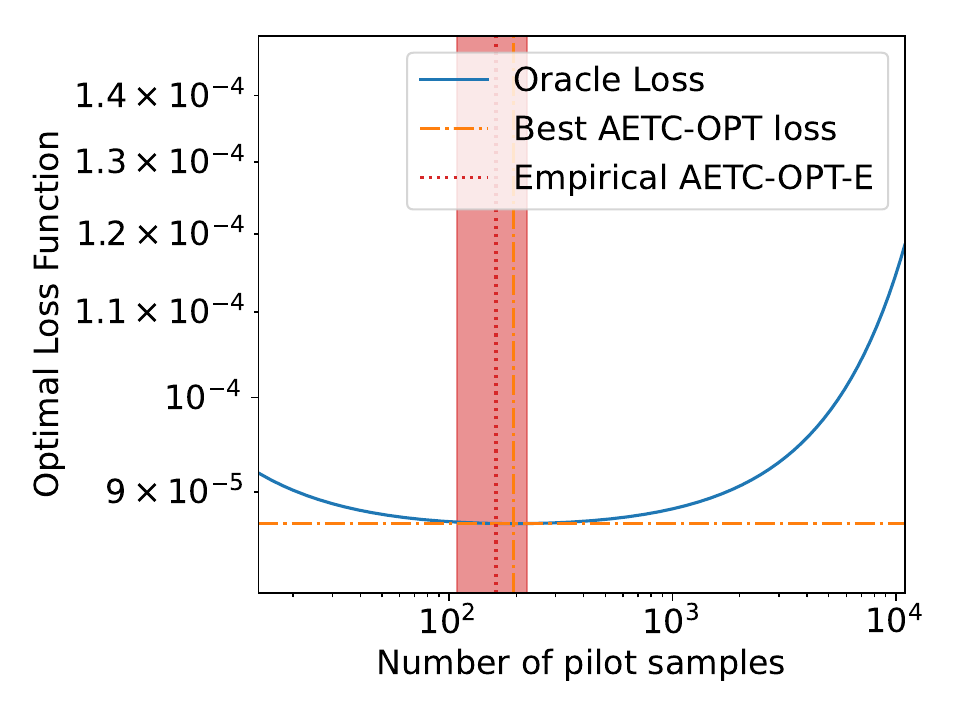}\hfill
\includegraphics[width=0.42\textwidth]{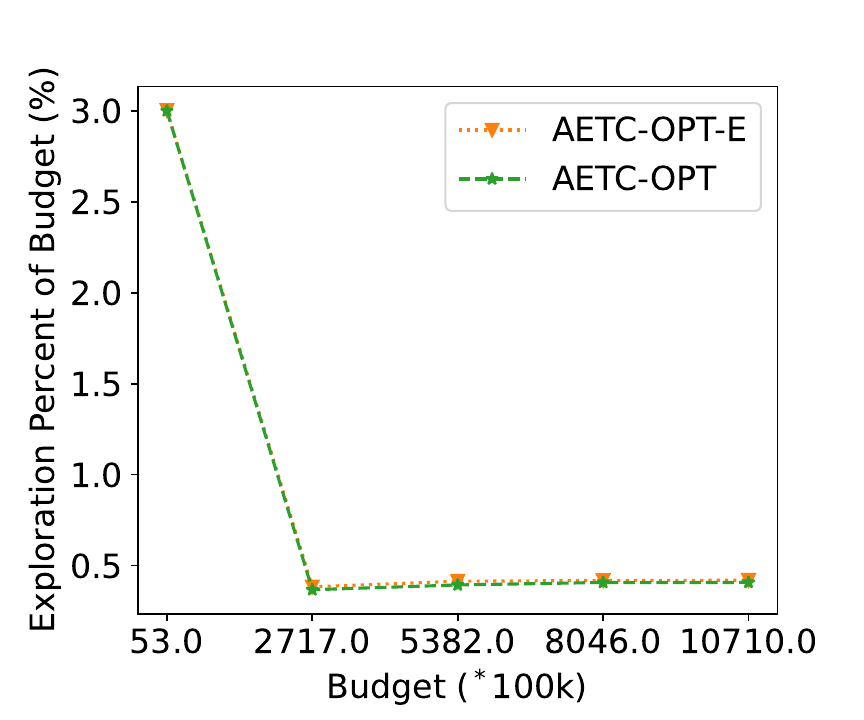}
\caption{\small
({\it Left}) The impact of the number of exploration samples on the MSE of the AETC-OPT algorithm. The AETC-OPT-E algorithm finds the optimal number of exploration samples, denoted by the red region. ({\it Right}) The percentage of the total budget given to exploration sampling in the AETC-OPT and AETC-OPT-E algorithms across the budgets.}
  \label{fig:ice-pilot}
  \end{center}
\end{figure}

The left panel of \Cref{fig:ice-pilot} plots the optimal loss curve across all subsets as a function of the number of exploration samples for the largest budget. As with the previous numerical example, discussed in \Cref{sec:elastic-pde}, the AETC-OPT-E algorithm can identify the optimal number of exploration samples. Additionally, as shown in the right panel of \Cref{fig:ice-pilot}, the percentage of the budget given to exploration sampling is remarkably low at 0.5\%. 

\begin{figure}[htbp]
\begin{center}
\includegraphics[width=0.32\textwidth,valign=t]{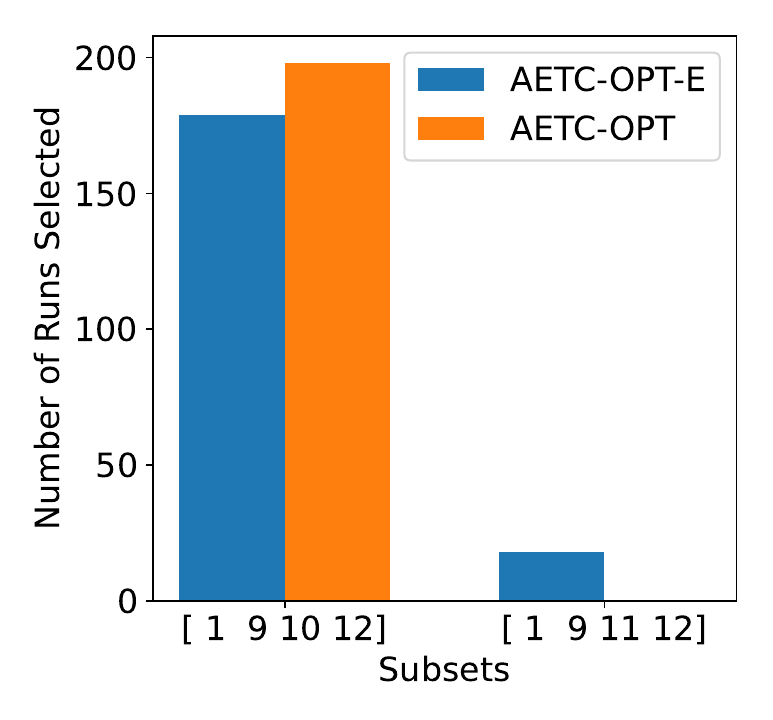}\hfill
\includegraphics[width=0.68\textwidth,valign=t]{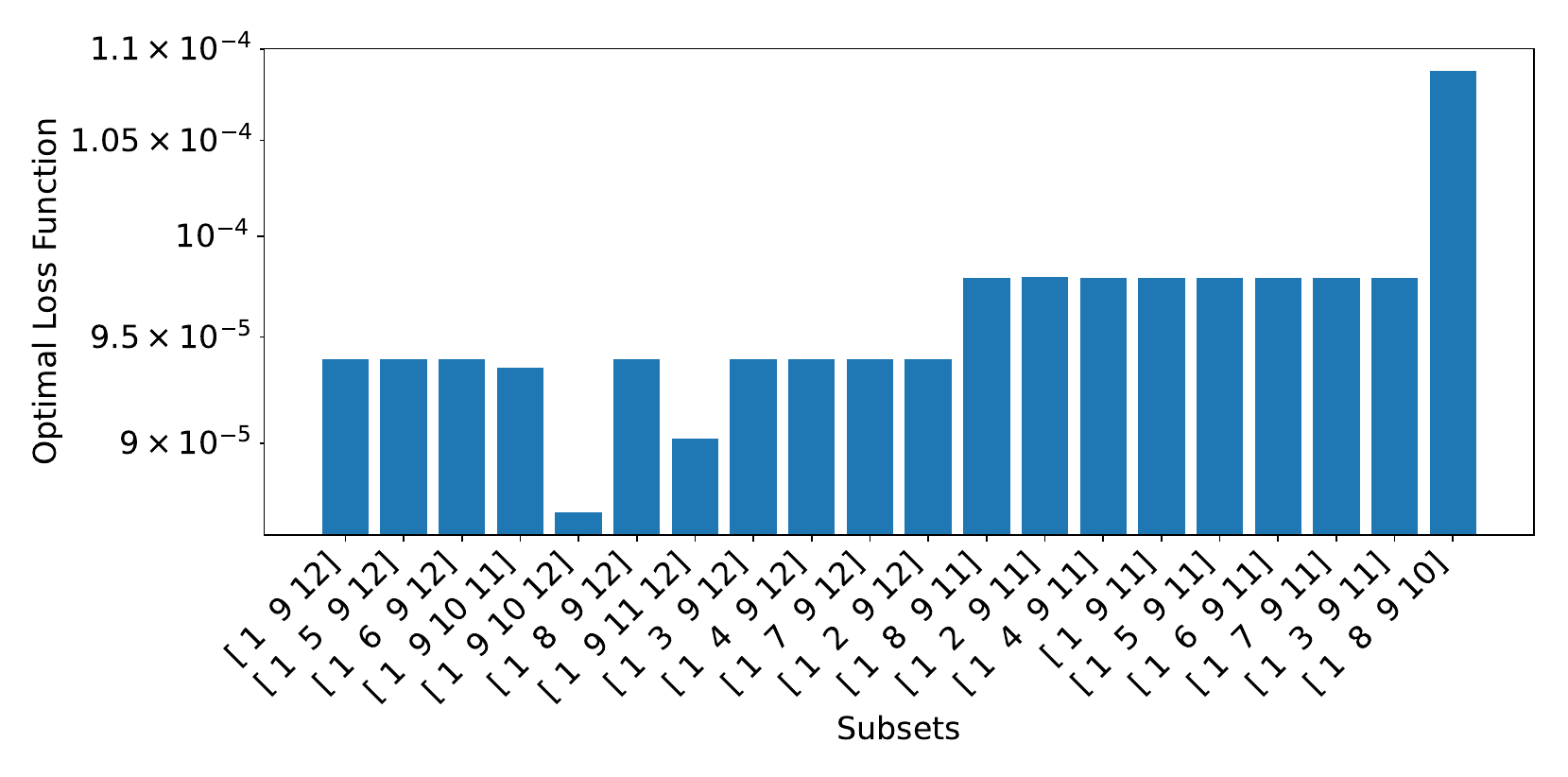}
\caption{\small ({\it Left}) The frequency of the exploitation subsets chosen by AETC-OPT and AETC-OPT-E when used to estimate the mean mass change. ({\it Right}) The best oracle MSE for each subset.
 }
  \label{fig:ice-subset}
  \end{center}
\end{figure}

The AETC-OPT algorithms not only optimally balance the cost of exploration and exploitation but also determine the model subset that minimizes the estimator variance. \Cref{fig:ice-subset} displays the selection of the model subsets for exploitation. Specifically, the left plot depicts the number of times the AETC-OPT and AETC-OPT-E algorithms selected specific subsets. The resulting LRMC estimators only chose 2 of the 20 subsets to exploit, corresponding to the 2 lowest oracle MSE, as shown on the right bar chart in \Cref{fig:ice-subset}. The majority of the AETC-OPT algorithms chose the subset with the smallest oracle MSE. This demonstrates the ability of the AETC-OPT algorithms to identify and exploit the correct subset from the available models.

The exact impact of balancing exploration and exploitation costs is problem-dependent. To illustrate this, we reduce the number of available models to force the AETC-OPT/AETC-OPT-E algorithms to use low-correlation models. We chose models 0, 3, 4, 7, 8, 11, and 12 such that the low-fidelity models with the highest correlations are not included. 
We set the regularization parameter $\alpha_q=2^{-q}$ to encourage higher exploration rates since the low-fidelity models have lower correlation with the high-fidelity model.
As before, the total budgets selected are chosen so that the highest budget uses most of the available exploration samples. To improve the visualization of results, we use AETC-OPT to iterate only over the 14 subsets with the lowest oracle MSE. 

\begin{figure}[htbp]
\begin{center}
\includegraphics[width=0.40\textwidth]{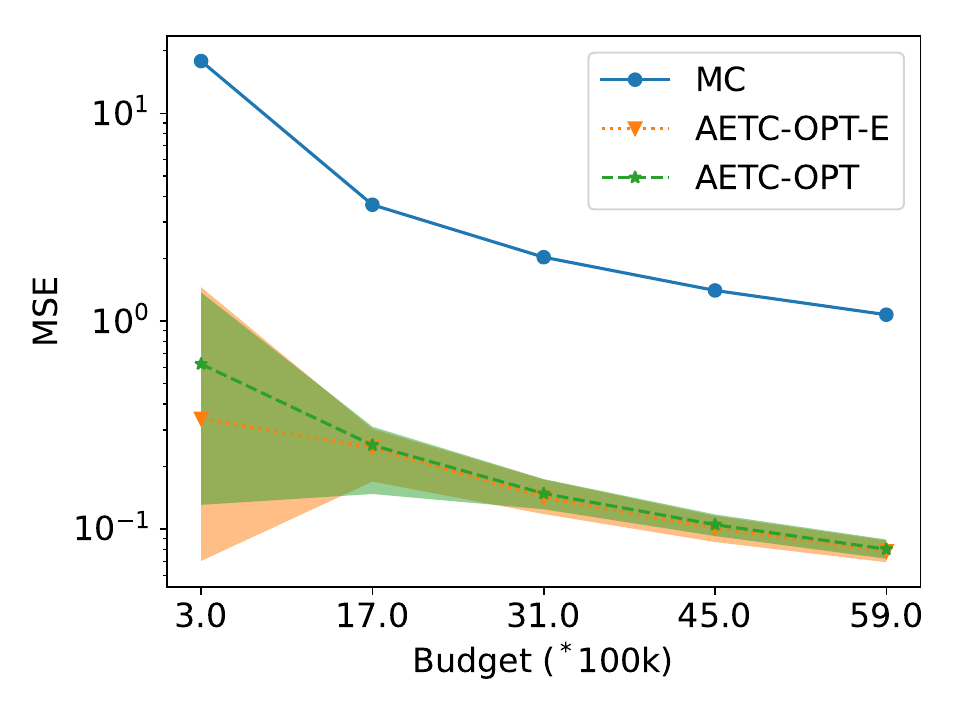}\hfill
\includegraphics[width=0.40\textwidth]{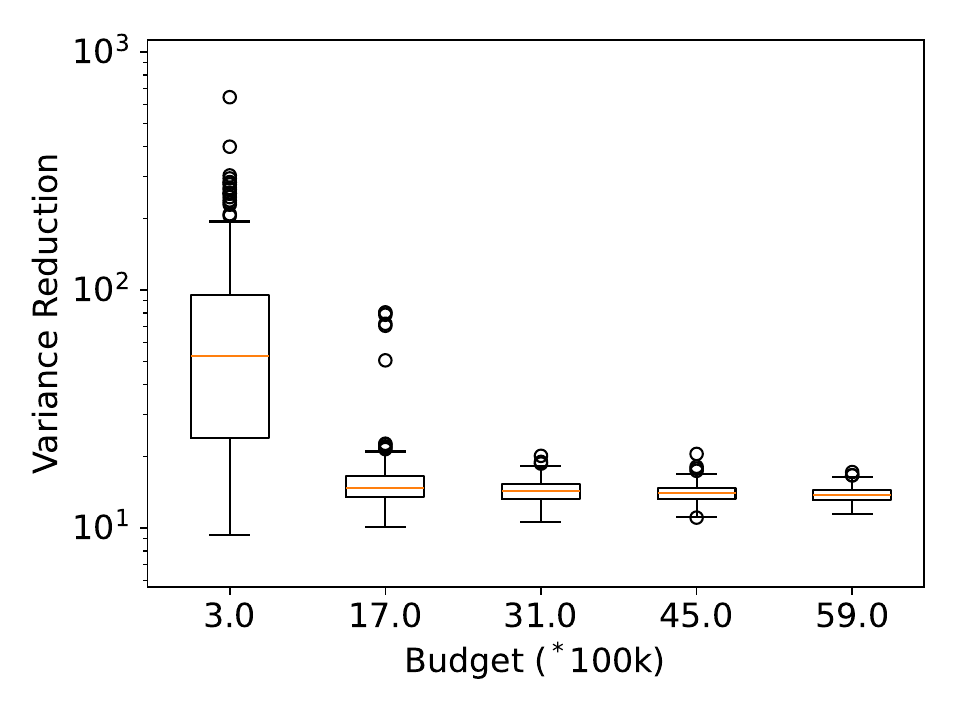}
\caption{\small {\it (Left)} Comparison of AETC-OPT-E with MC when used to estimate the mean mass loss from Humboldt-Glacier with seven models. The dotted line represents the median loss, and the edges of the shaded region represent the 0.05-0.95 quantiles. {\it (Right)} The variance reduction of using the AETC-OPT-E algorithm for increasing budgets. 
 }
  \label{fig:lf-ice-budget}
  \end{center}
\end{figure}

The left panel of \Cref{fig:lf-ice-budget} plots the MSE of the LRMC estimators produced by AETC-OPT and AETC-OPT-E. The median loss of AETC-OPT-E is very similar to that of AETC-OPT, despite the former not using oracle statistics. However, the loss of both methods has increased relative to the loss reported in \Cref{fig:ice-budget} due to the forced use of poorly correlated models. Compared to \Cref{fig:ice-budget}, the quantiles are larger, validating our observation that low-correlation models typically require a larger number of exploration samples. 
The poor correlation between models in the ensemble also results in a smaller variance reduction of 14.0, as shown in the right plot of \Cref{fig:lf-ice-budget}.

\begin{figure}[htbp]
\begin{center}
\includegraphics[width=0.40\textwidth]{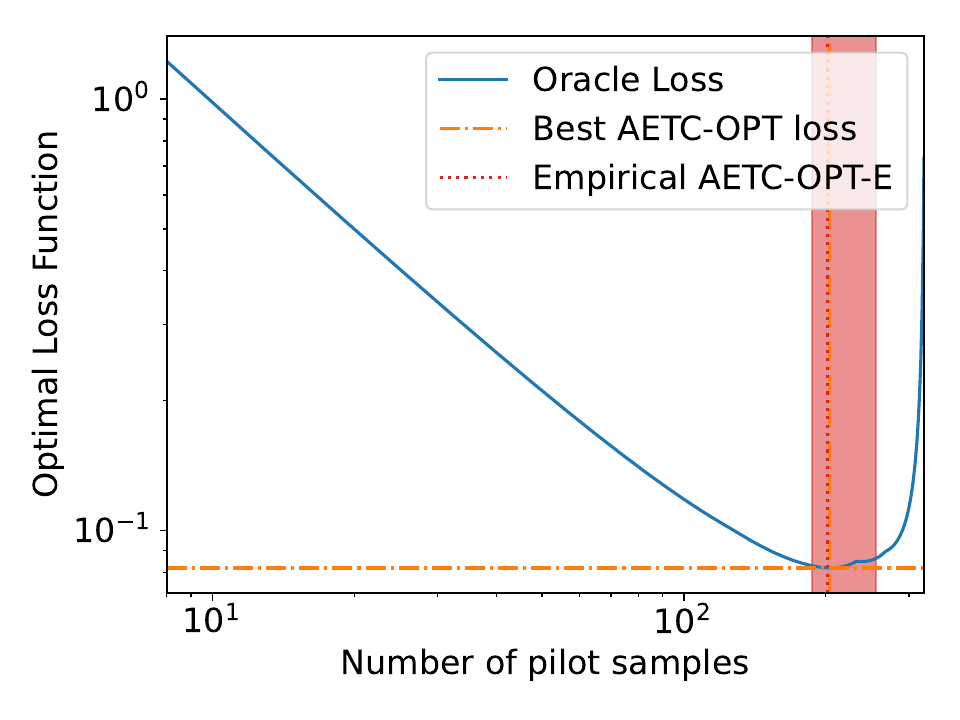}\hfill
\includegraphics[width=0.40\textwidth]{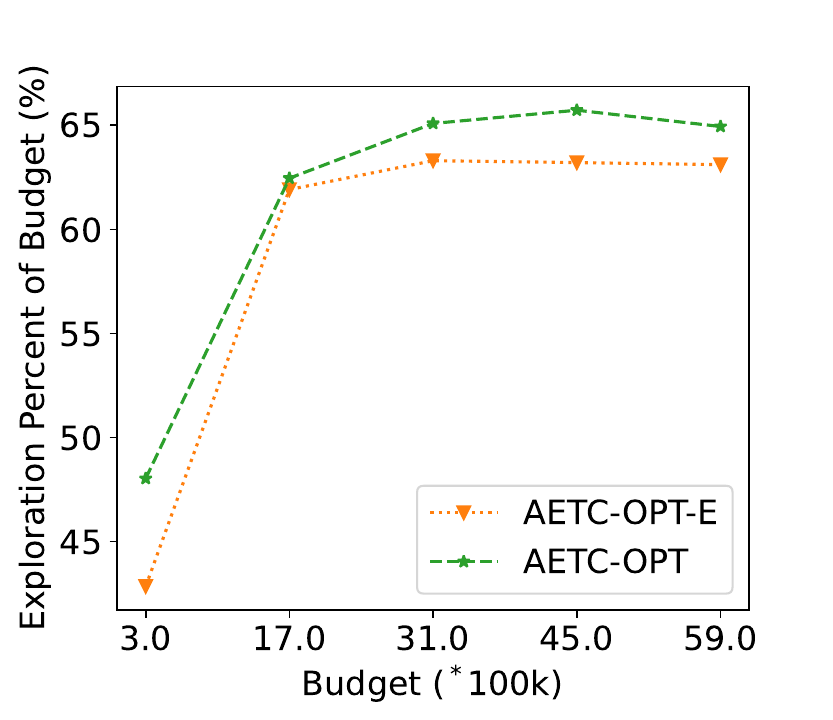}
\caption{\small ({\it Left}) The impact of the number of exploration samples on the MSE of mass-change when using seven models.
({\it Right}) The percentage of the total budget given to exploration sampling in the AETC-OPT and AETC-OPT-E algorithms across the budgets.
 }
  \label{fig:lf-ice-pilot}
  \end{center}
\end{figure}

More interestingly, \Cref{fig:lf-ice-pilot} displays the optimal loss curve as a function of the number of exploration samples determined by the AETC-OPT algorithm, which is now a much larger percentage of the total budget. This demonstrates that the lack of available high-correlation models causes the AETC-OPT algorithm to require a more extensive exploration phase. The optimal loss curve on the left signals that choosing the correct number of exploration samples is essential for minimizing the MSE of a multi-fidelity estimator. We notice a trend that implies when high-correlation low-fidelity models are available, fewer exploration samples are needed, but when there are no high-correlation low-fidelity models, more exploration samples are required to estimate the necessary exploration statistics.

\begin{figure}[htbp]
\begin{center}
\includegraphics[width=0.4\textwidth,valign=t]{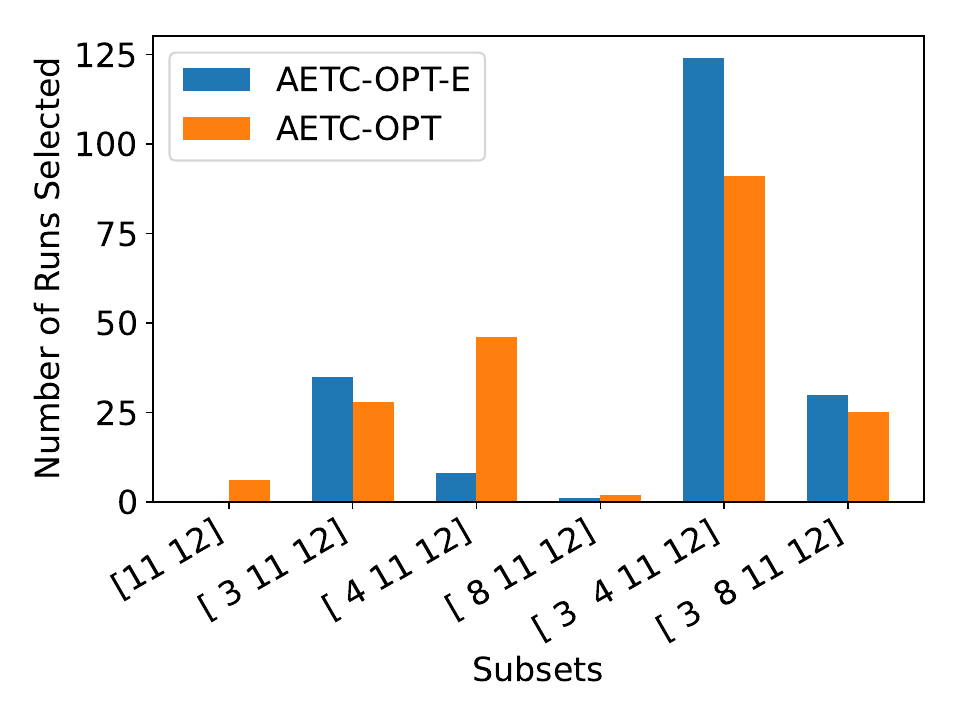}\hfill
\includegraphics[width=0.58\textwidth,valign=t]{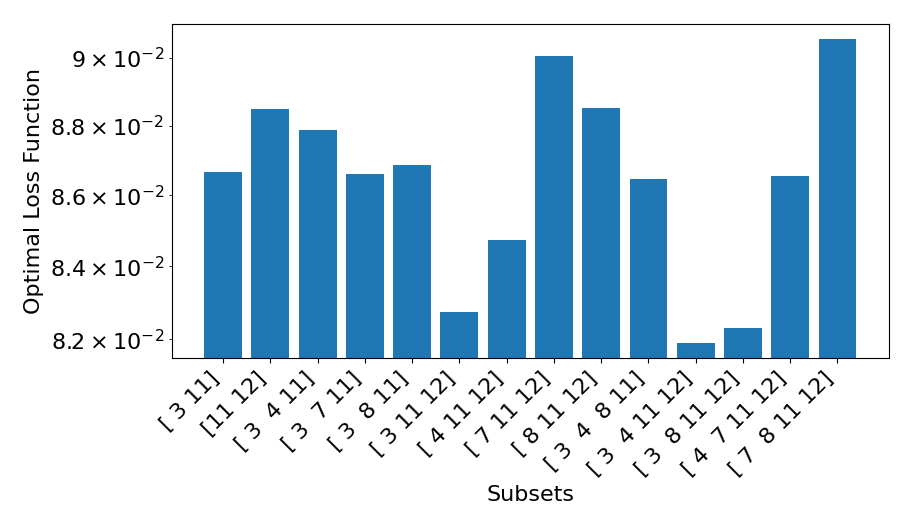}
\caption{\small ({\it Left}) The frequency of the exploitation subsets chosen by AETC-OPT and AETC-OPT-E when used to estimate the mean mass change. ({\it Right}) The best oracle MSE for each subset.
 }
  \label{fig:lf-ice-subset}
  \end{center}
\end{figure}

Lastly, since the exploration phase is much larger than the previous results, we see that more subsets were chosen in \Cref{fig:lf-ice-subset}. This result implies that there is more volatility in the exploration phase and that more exploration samples are needed to converge to the correct optimal subset than when using all 13 models. Despite this, most of the estimators chose the subset with the lowest oracle MSE. These results demonstrate the capability of the AETC-OPT algorithm to identify the optimal number of exploration samples and optimal subset, which can vary widely and is extremely problem-dependent.

%

\section{Conclusion}

In this work, we build on the bandit-learning framework in \cite{xu2022bandit} to introduce new estimation methods. By generalizing the previously introduced bandit-learning scheme, we propose more efficient estimation methods in the exploitation phase of the AETC algorithm based on MLBLUEs with optimized sample allocations to improve the MSE of the adaptively constructed multi-fidelity estimator. We establish the convergence of the proposed multi-fidelity estimator and other desirable properties such as optimality and robustness. By applying MLBLUEs within the bandit-learning framework, we improve the MSE of the resulting multi-fidelity estimator for the high-fidelity mean to a level that nearly matches the optimal lower bound. We demonstrate the autonomy of the generalized AETC algorithm (i.e., AETC-OPT) in identifying the optimal number of exploration samples and the optimal subset in a variety of applications. 

The cost of taking exploration samples to estimate the required statistics should not be ignored when applying multi-fidelity estimation techniques. The results presented in this work highlight the importance of determining the optimal number of exploration samples, especially since it is highly problem-dependent. Future work will extend the bandit-learning scheme to other estimation tasks, such as variance or sensitivity indices. 

\section*{Acknowledgment}
We would like to thank the Editors, the Associate Editors, and the anonymous referees for their constructive comments, which significantly improved the manuscript’s presentation. Dixon and Jakeman were supported by the US Department of Energy’s Office of Advanced Scientific Computing Research program. Gorodetsky was supported by the National Nuclear Security Administration’s Advanced Simulation and Computing (ASC) program. Narayan was supported by AFOSR FA9550-20-1-0338 and AFOSR FA9550-23-1-0749. Xu was supported by the University of Kentucky for the start-up funding and the AMS-Simons Travel grant (No.3048116562). 

Sandia National Laboratories is a multi-mission laboratory managed and operated by National Technology \& Engineering Solutions of Sandia, LLC (NTESS), a wholly owned subsidiary of Honeywell International Inc., for the U.S. Department of Energy’s National Nuclear Security Administration (DOE/NNSA) under contract DE-NA0003525. This written work is authored by an employee of NTESS. The employee, not NTESS, owns the right, title, and interest in and to the written work and is responsible for its contents. Any subjective views or opinions that might be expressed in the written work do not necessarily represent the views of the U.S. Government. The publisher acknowledges that the U.S. Government retains a non-exclusive, paid-up, irrevocable, worldwide license to publish or reproduce the published form of this written work or allow others to do so, for U.S. Government purposes. The DOE will provide public access to the results of federally sponsored research in accordance with the DOE Public Access Plan.

\bibliographystyle{siamplain}

\bibliography{ref}

\newpage

\begin{appendix}
\section{Variance of Independent Random Vectors} \label{ap:independent_variance}
Let $X$ and $Y$ be independent random vectors with bounded second moments and compatible sizes. The variance of their inner product $X^\top Y$ can be computed as 
\begin{align*}
    \V[X^\top Y]=&\ \E[(X^\top Y)^2] - \E[X^\top Y]^2 \nonumber\\
    =&\ \E[\Tr(X^\top YY^\top X)] - (\E[X]^\top\E[Y])^2 \nonumber\\
    =&\  \Tr(\E[YY^\top]\E[XX^\top]) - (\E[X]^\top\E[Y])^2 \nonumber\\
    =&\ \Tr((\Cov[Y] + \E[Y]\E[Y]^\top)(\Cov[X] + \E[X]\E[X]^\top)) - (\E[X]^\top\E[Y])^2 \nonumber\\
    =&\  \Tr(\Cov[Y]\Cov[X] + \E[Y]\E[Y]^\top\Cov[X]  + \Cov[Y]\E[X]\E[X]^\top) \nonumber\\
    =&\  \Tr(\Cov[Y]\Cov[X]) + \E[Y]^\top\Cov[X]\E[Y]  + \E[X]^\top\Cov[Y]\E[X].
\end{align*}

\section{Proof of \Cref{lemma:bl-asymptotic-mse}}\label{sec:supp2}
Let $\E_W [\cdot]$ and $\E_{\eta_S}[\cdot]$ represent the expectation over the exploitation samples and exploration discrepancy, respectively. It follows from the tower property of conditional expectations and the law of large numbers that, a.s.,  
\begin{align}
& \mse_S  = \E_{\eta_S,W}\left[\left(\widehat{y}^\top _S\widehat{\beta}_S-\mu_0\right)^2\bigg |~Z_\lfs\right]\label{minger} \\
\stackrel{\eqref{lr}}{=}&\ \E_{\eta_S,W}\left[\left(\widehat{y}^\top _S\widehat{\beta}_S-y_S^\top\beta_S\right)^2\bigg |~Z_\lfs\right]\nonumber\\
 =&\ \E_{\eta_S}\left[\E_{W}\left[\left(\widehat{y}^\top _S\widehat{\beta}_S-y_S^\top \beta_S\right)^2\bigg |~Z_\lfs, \widehat{\beta}_S\right]\bigg|~Z_{[n]}\right]&\text{(tower property)}\nonumber\\
=&\ \E_{\eta_S}\left[\left(y^\top_S\widehat{\beta}_S  - y_S^\top \beta_S\right)^2 \bigg |~Z_\lfs\right] + \E_{\eta_S}\left[\widehat{\beta}_S^\top \Cov_W[\widehat{y}_S\mid\widehat{\beta}_S, Z_\lfs]\widehat{\beta}_S \bigg |~Z_\lfs\right]&\text{(exploration-unbiasedness)}\nonumber\\
 \stackrel{\eqref{lse}}{=}&\ \E_{\eta_S}\left[\left(y^\top_S Z_S^\dagger\eta_S\right)^2 \bigg |~Z_\lfs\right] + \E_{\eta_S}\left[\widehat{b}^\top_S\Cov_W[\widehat{\mu}_S\mid\widehat{b}_S,Z_\lfs]\widehat{b}_S \bigg |~Z_\lfs\right]\nonumber\\
\stackrel{\eqref{gamma}}{\simeq}&\ \E_{\eta_S}\left[\left(y^\top _S Z_S^\dagger\eta_S\right)\left(y^\top_S Z_S^\dagger\eta_S\right)^\top \bigg |~Z_\lfs\right] + \frac{\gamma(S)}{B - c_\ex q} \nonumber\\
=&\ \tr\left(y_S  y^\top _S Z_S^\dagger \E_{\eta_S}\left[\eta_S \eta_S^\top \bigg |~Z_\lfs\right] Z_S^{\dagger,\top}\right)  + \frac{\gamma(S)}{B - c_\ex q}  \nonumber\\
=&\ \sigma_S ^2\tr\left(y_S  y^\top_S Z_S^\dagger Z_S^{\dagger,\top}\right)  + \frac{\gamma(S)}{B - c_\ex q} &\text{($\E[\eta_S\eta_S^\top] = \sigma_S^2 I_q$)} \nonumber\\
=&\ \sigma_S^2\tr\left(y_S  y^\top _S (Z_S^\top Z_S)^{-1}\right)  + \frac{\gamma(S)}{B - c_\ex q}  \nonumber\\
\simeq&\ \frac{\tr(y_S y_S ^\top \Pi_S ^{-1})\sigma_S^2}{q} + \frac{\gamma(S)}{B - c_\ex q}&\text{($Z_S^\top Z_S/q\to\Pi_S$)}\nonumber\\
\stackrel{\eqref{newobs}}{=}&\ \frac{\sigma_S^2}{q} + \frac{\gamma(S)}{B - c_\ex q}.\nonumber 
\end{align}

\section{Proofs in \Cref{sec:4.1}}\label{sec:supp4.1}
\subsection{Proof of \Cref{4210}}\label{sec:supp3}
Note that $Z_\lfs$ and $\widehat{b}_S$ affect the sample allocation $\mathcal M$ in exploitation. Thus, $\widehat{\mu}_{S}$ is not independent of the exploration data. However, conditional on $\mathcal M$, the MLBLUE $\widehat{\mu}_S$ for $\mu_S$ uses exploitation samples only and thus is unbiased for $\mu_S$. Thanks to the independence between exploration and exploitation samples, this result holds regardless of whether $\Sigma_S$ or $\widehat{\Sigma}_S$ is used (see also \eqref{appunb}). Hence, $\E[\widehat{\mu}_S\mid \widehat{b}_S,Z_\lfs] = \mu_S$, as desired. 

\subsection{Proof of \Cref{4211}}\label{sec:supp4}
Since we are interested in the case when $B_\ext\to\infty$, we consider \eqref{optint} with $\afs$ replaced by $S$ and assume $\mathcal M = \{m_T\}_{T\subseteq S}\in\R_+^{2^{s}-1}$ is continuous. The difference between the optimal continuous and discrete allocations is asymptotically negligible as $B_\ext\to\infty$. Moreover, we assume that $\widehat{\mu}_S$ is computed using the empirical covariance estimator $\widehat{\Sigma}_S$ from the exploration phase, as the proof for the oracle case is almost identical. 

Let $J({\mathcal M}_\opt(B, a, \Sigma_S), a, \Sigma_S)$ denote the optimal objective to \eqref{optint} with $\afs$ replaced by $S$, where ${\mathcal M}_\opt(B, a, \Sigma_S)\in\R_+^{2^{s}-1}$ is the optimal continuous sample allocation when supplemented with budget $B$, sketch $a$, and covariance $\Sigma_S$.  In particular, $J({\mathcal M}_\opt(B, a, \Sigma_S), a, \Sigma_S)=\var[a^\top\widehat{\mu}_S(B, a, \Sigma_S)]$, where $\widehat{\mu}_S(B, a, \Sigma_S)$ is the MLBLUE under ${\mathcal M}_\opt(B, a, \Sigma_S)$. Using this notation, we have 
\begin{align*} 
J({\mathcal M}_{\opt}(B_\ext, \widehat{b}_S, \widehat{\Sigma}_S), \widehat{b}_S, \widehat{\Sigma}_S) = \widehat{b}_S^\top \Cov[\widehat{\mu}_S\mid \widehat{b}_S,Z_\lfs]\widehat{b}_S, 
\end{align*}
where we have used $\widehat{\mu}_S = \widehat{\mu}_S(B_\ext, \widehat{b}_S, \widehat{\Sigma}_S)$ (which takes $\widehat{b}_S$ and $\widehat{\Sigma}_S$ as input when finding the optimal MLBLUE). Since $J$ scales inversely linearly in $1/B$, 
\begin{align*}
J({\mathcal M}_{\opt}(1, \widehat{b}_S, \widehat{\Sigma}_S), \widehat{b}_S, \widehat{\Sigma}_S) = B_\ext\cdot \widehat{b}_S^\top \Cov[\widehat{\mu}_S\mid\widehat{b}_S, Z_\lfs]\widehat{b}_S.
\end{align*}
Therefore, it suffices to show 
\begin{align*}
&\lim_{q\to\infty}\E[J({\mathcal M}_{\opt}(1, \widehat{b}_S, \widehat{\Sigma}_S), \widehat{b}_S, \widehat{\Sigma}_S)\mid Z_\lfs]=\gamma(S)& a.s.
\end{align*}
for some constant $\gamma(S)\in (0, \infty)$. 

To this end, note under \Cref{subexp}, by standard moment bounds, the following events hold with probability at least $1-q^{-3/2}$ for all large $q$: 
\begin{align}
\left\|\frac{1}{q}Z_S^\top Z_S - \Pi_S\right\|_F\lesssim q^{-1/10}\Rightarrow \tr((Z_S^\top Z_S)^{-1}) \lesssim q^{-1}, \quad\quad \left\|\widehat{\Sigma}_S - \Sigma_S\right\|_F\lesssim q^{-1/10}, \label{lo989}
\end{align}
where $\Pi_S$ is defined in \eqref{snota}. 
By the Borel--Cantelli lemma, the events in \eqref{lo989} hold for all large $q$ a.s. 

In the subsequent analysis, we condition on $Z_\lfs$ and assume the events in \eqref{lo989} hold.   
Under \Cref{subgaussian}, $\widehat{b}_S$ is close to $b_S$ with high probability by the Hanson--Wright inequality \cite[Theorem 6.2.1]{Vershynin2018}:  
\begin{align}
\P\left(\|\widehat{b}_S-  b_S\|^2_2\lesssim \E[\|\widehat{b}_S-  b_S\|^2_2\mid Z_\lfs]\log q\right)\geq 1-q^{-2}.\label{6612}
\end{align}
Under the linear model assumption~\eqref{lr}, 
\begin{align}
\E[\|\widehat{b}_S-  b_S\|^2_2\mid Z_\lfs] = \sigma^2_S\tr((Z_S^\top Z_S)^{-1})\stackrel{\eqref{lo989}}{\lesssim}\frac{\sigma_S^2}{q}.\label{tdfg}
\end{align}
The event in \eqref{6612} thus implies that there exists $C>0$ such that  
\begin{align}
\|\widehat{b}_S-  b_S\|^2_2\leq \frac{C\sigma_S^2\log q}{q}\Longrightarrow \|\widehat{b}_S-  b_S\|_2\leq q^{-1/3}, 
\end{align}
Consequently, 
\begin{align*}
&\left|\E[J({\mathcal M}_{\opt}(1, \widehat{b}_S, \widehat{\Sigma}_S), \widehat{b}_S, \widehat{\Sigma}_S)\mid Z_\lfs] - J({\mathcal M}_{\opt}(1, b_S, \Sigma_S), b_S, \Sigma_S)\right|\\
\leq&\ \E[|J({\mathcal M}_{\opt}(1, \widehat{b}_S, \widehat{\Sigma}_S), \widehat{b}_S, \widehat{\Sigma}_S)-J({\mathcal M}_{\opt}(1, b_S, \Sigma_S), b_S, \Sigma_S)|\cdot \mathbf 1_{\{\|\widehat{b}_S-  b_S\|_2\leq q^{-1/3}\}}\mid Z_\lfs]\ + \\
& \E[|J({\mathcal M}_{\opt}(1, \widehat{b}_S, \widehat{\Sigma}_S), \widehat{b}_S, \widehat{\Sigma}_S)-J({\mathcal M}_{\opt}(1, b_S, \Sigma_S), b_S, \Sigma_S)|\cdot \mathbf 1_{\{\|\widehat{b}_S-  b_S\|_2> q^{-1/3}\}}\mid Z_\lfs]\\
\stackrel{\eqref{lo989}}{\lesssim} &\ \max_{\substack{\|a-b_S\|_2\leq q^{-1/3}\\ \|\Gamma_S-\Sigma_S\|_2\leq q^{-1/10}}}|J({\mathcal M}_{\opt}(1, b_S, \Sigma_S), b_S, \Sigma_S)-J({\mathcal M}_{\opt}(1, a, \Gamma_S), a, \Gamma_S)| \ + \\
& \E[(J({\mathcal M}_{\opt}(1, \widehat{b}_S, \widehat{\Sigma}_S), \widehat{b}_S, \widehat{\Sigma}_S)+J({\mathcal M}_{\opt}(1, b_S, \Sigma_S), b_S, \Sigma_S))\cdot \mathbf 1_{\{\|\widehat{b}_S-  b_S\|_2> q^{-1/3}\}}\mid Z_\lfs]. 
\end{align*}
When $q\to\infty$, the first term goes to zero because of the continuity assumption in \Cref{myass}. For the second term, one can always use the uniform continuous allocation under a unit budget (i.e., $m_S = 1/c_S$ and $m_T = 0$ for $T\neq S$) to obtain an upper bound on $J$, 
\begin{align}
J({\mathcal M}_{\opt}(1, \widehat{b}_S, \widehat{\Sigma}_S), \widehat{b}_S, \widehat{\Sigma}_S)\leq c_S\widehat{b}_S^\top\Sigma_S\widehat{b}_S\leq c_S\|\widehat{b}_S\|_2^2\|\Sigma_S\|_2.\label{pokil}
\end{align}
As a result, 
\begin{align*}
&\E[(J({\mathcal M}_{\opt}(1, \widehat{b}_S, \widehat{\Sigma}_S), \widehat{b}_S, \widehat{\Sigma}_S)+J({\mathcal M}_{\opt}(1, b_S, \Sigma_S), b_S, \Sigma_S))\cdot\mathbf 1_{\{\|\widehat{b}_S-  b_S\|_2> q^{-1/3}\}}\mid Z_\lfs]\\
\stackrel{\eqref{pokil}}{\leq} &\  c_S\|\Sigma_S\|_2\cdot\E[(\|\widehat{b}_S\|_2^2+\|b_S\|_2^2)\cdot \mathbf 1_{\{\|\widehat{b}_S-  b_S\|_2> q^{-1/3}\}}\mid Z_\lfs]\\
\stackrel{\eqref{6612}}{\leq}&\ 2c_S\|\Sigma_S\|_2\left(\E[\|\widehat{b}_S-b_S\|_2^2\cdot \mathbf 1_{\{\|\widehat{b}_S-  b_S\|_2> q^{-1/3}\}}\mid Z_\lfs] + \frac{3\|b_S\|_2^2}{q^2}\right)\\
=&\ 2\|\Sigma_S\|_2\left(\E[q\|\widehat{b}_S-b_S\|_2^2\cdot \mathbf 1_{\{q\|\widehat{b}_S-  b_S\|^2_2> q^{1/3}\}}\mid Z_\lfs]  + \frac{3\|b_S\|_2^2}{q^2}\right)\xrightarrow{q\to\infty}{0},
\end{align*}
where the last inequality follows by noting $\E[q\|\widehat{b}_S-  b_S\|^2_2\mid Z_\lfs]\lesssim \sigma_S^2\lesssim 1$ because of \eqref{tdfg}. Hence, the desired result follows by taking $\gamma(S) = J({\mathcal M}_{\opt}(1, b_S, \Sigma_S), b_S, \Sigma_S)$. 

\subsection{Proof of \Cref{4212}}\label{sec:supp5}

Assumptions \ref{subgaussian}-\ref{subexp} ensure that $\widehat{b}_S\to b_S$, $\widehat{\Sigma}_\lfs \to\Sigma_\lfs$, and $\widehat{c}_\lfs\to c_\lfs$ as $q\to\infty$ a.s. thanks to the law of large numbers (i.e., $\widehat{\Sigma}_T\to\Sigma_T$ a.s. for all $T\subseteq \lfs$). The desired result follows immediately due to \Cref{myass}.

\section{Proofs in \Cref{sec:alga}}\label{sec:supp4.2}

\subsection{Proof of \Cref{thm:AETC}}\label{sec:supp6}

Our proof is adapted from \cite[Theorem 5.3]{xu2021budget}, which employs a similar bisection-accelerated bandit-learning procedure for distribution estimation rather than computation of expectations. The major difference here is the additional randomness of costs, which requires a few modifications.  

Before beginning the proof, we introduce some notation that will be used in the analysis. Specifically, we let $q$ denote the exploration rate as in the main section that grows non-linearly with respect to an index that counts the iterations of the loop in step 3 of \Cref{alg2}. Additionally, we let $p$ denote the loop iteration index, and $q_p$ the corresponding exploration rate, i.e., $q_1 = n+2$. Then, letting $p(B)$ be the total iteration steps in \Cref{alg2}, which is random, it follows from the definition that $q_{p(B)} = q(B)$ and 
\begin{align}
&n+2\leq q_p\leq q_{p+1}\leq 2q_p& 1\leq p< p(B),\label{myq}
\end{align} 
which can be seen from the bisection update in steps 10-14 of \Cref{alg2}. 
Morever, since the random costs $\{\C_i\}_{i\in\afs}$ are assumed to be uniformly bounded, we let $c^* := \max_{i\in\afs}\C_i<\infty$ denote its maximum value (i.e., $c_\ex\leq (n+1)c^*$). We complete the proof in two steps: first, we show that $q(B)$ diverges a.s.; then, based on this, we argue the asymptotic optimality of the selected model and exploration rate.

\smallskip

\underline{\textbf{Step 1}}: \textbf{Divergence of $q(B)$ a.s.}


According to \Cref{4212} and the assumption on $\widehat{k}_q(S)$, $\widehat{\gamma}_q(S)\to\gamma(S)$ and $0<\widehat{k}_q(S)\to k(S)$ for all $S\subseteq [n]$ as $q\to\infty$ a.s. As a result, for almost every realization $\omega\in\Omega$ (where $\Omega$ denotes the product space of exploration samples and corresponding costs), there exists an $L(\omega)<\infty$ such that $\sup_{q>n+1}\max_{S\subseteq [n]}\widehat{\gamma}_q(S; \omega)<L(\omega)<\infty$. 
Because of the randomness in costs, there are two potential scenarios when the algorithm terminates:
\begin{enumerate}
\item [(i)] The algorithm proceeds to exploitation through lines 23-24;
\item [(ii)] Either condition in line 9 or line 16 is met, but the sample collection try fails due to running out of remaining budget (only possible when costs are random).
\end{enumerate}
In case (i), the exploration stopping criteria of \Cref{alg2} require that 
\begin{align}
q(B; \omega)\geq \widehat{q}^*_{S(B; \omega)}(q(B; \omega); \omega)&\stackrel{\eqref{getq*}}{=} \frac{B}{
  \widehat{c}_{\ex}
  + \sqrt{ 
      \frac{
        \widehat{c}_{\ex}\,\widehat{\gamma}_{q(B; \omega)}(S(B;\omega))
      }{
        \widehat{k}_{q(B; \omega)}(S(B;\omega))
      }
    }
}\nonumber\\
& \geq \frac{B}{(n+1)c^*+\sqrt{\frac{(n+1)c^* L(\omega)}{\alpha_q}}}, \label{87421}
\end{align} 
where the lower bound diverges as $B\to\infty$ if $q(B; \omega)$ is uniformly bounded. 
In case (ii), since the cost of each exploration sample is at most $(n+1)c^*$, the total exploration cost is upper bounded by $(n+1)c^*q(B; \omega)$. By the stopping criteria, 
\begin{align}
(n+1)c^*q(B; \omega)\geq B\Longrightarrow q(B; \omega)\geq\frac{B}{(n+1)c^*}.\label{87422} 
\end{align}
Combining \eqref{87421}-\eqref{87422} concludes that 
\begin{align}
&\lim_{B\to\infty}q(B; \omega)=\infty. 
\label{ksl1}
\end{align}

\smallskip

\underline{\textbf{Step 2}}: \textbf{Asymptotic optimality of model selection and exploration rate}


Using the strong law of large numbers, one can verify that all estimators in \eqref{haozi} converge to the true parameters a.s. In particular, $\widehat{k}_{q}(S)$ and $\widehat{\gamma}_q(S)$ are consistent as $q\to\infty$. Now fix a realization $\omega$ such that $q(B; \omega)\to\infty$ as $B\to\infty$, and all estimators in \eqref{haozi} converge to the true parameters as $q\to\infty$. To prove that both \eqref{rr1} and \eqref{rr2} hold for such an $\omega$, we first consider case (i) and then exclude the possibility of case (ii). 

We first consider the scenario when case (i) holds. Fix $\delta<\min\{1/2, \sqrt{\gamma(S^*)/c_\ex k(S^*)}/4\}$ sufficiently small. Since $S^*$ is assumed unique, a continuity argument implies that there exists a sufficiently large $T(\delta; \omega)$, such that for all $q\geq T(\delta; \omega)$, 
\begin{align}
(1-\delta)c_\ex&\leq\widehat{c}_\ex(q)\leq (1+\delta) c_\ex\label{1113};\\
\max_{(1-\delta)q^*_{S^*}\leq z\leq (1+\delta)q^*_{S^*}}{\widehat{\L}}_{S^*}(z; q)&< \min_{S\subseteq [n], S\neq S^*}\widehat{\L}^*_S(q);\label{112}\\
1-\delta&\leq \frac{\widehat{q}^*_S(q; \omega)}{q^*_S}\leq 1+\delta&\forall S\subseteq [n]\label{113}.
\end{align}
To see why \eqref{112} holds for sufficiently small $\delta$, we take $B=1$ since otherwise one considers $B\L_{S^*}(q/B)$ instead. Since $S^*$ is unique, the maximum of $\L_{S^*}(q)$ in a small compact neighborhood of its minimizer, $q^*_{S^*}$, is strictly bounded by $\min_{S\subseteq [n], S\neq S^*}\L^*_S$. On the other hand, when $q\to\infty$, ${\widehat{\L}}_{S^*}(z; q)$ converges to $\L_{S^*}(z)$ uniformly in any compact neighborhood around $q_{S^*}$ and $\min_{S\subseteq [n], S\neq S^*}\L^*_S(q)\to\min_{S\subseteq [n], S\neq S^*}\L^*_S$ as a result of strong consistency. Thus, there exists a small $\delta$ such that \eqref{112} holds. 

Since $q^*_S$ scales linearly in $B$ and $q(B;\omega)$ diverges as $B\to\infty$, there exists a sufficiently large $B(\delta;\omega)$ such that for $B>B(\delta;\omega)$, 
\begin{align}
\min_{S\subseteq [n]}q^*_S&> 4T(\delta;\omega),\label{2345}\\
q_{p(B)} = q(B;\omega) &> 4T(\delta;\omega).\label{3456}
\end{align}
  Consider $p'< p(B)$ that satisfies $q_{p'-1}< T(\delta;\omega) \leq q_{p'}$. Such a $p'$ always exists due to \eqref{3456}, and satisfies 
\begin{align*}
q_{p'}\stackrel{\eqref{myq}}{\leq} 2q_{p'-1}<2T(\delta;\omega)\stackrel{\eqref{2345}}{\leq}\frac{1}{2}\min_{S\subseteq [n]}q^*_S\stackrel{\eqref{113}, \delta<1/2}{\leq}\widehat{q}^*_S(q_{p'}; \omega).
\end{align*} 
This inequality tells us that in the $p'$th loop iteration, for all $S\subseteq [n]$, the corresponding estimated optimal exploration rate is larger than the current exploration rate. 
In this case, 
\begin{align*}
&\mathcal R_S(q_{p'}) = \widehat{\L}_S(q_{p'}\vee \widehat{q}^*_S(q_{p'}; \omega); q_{p'}) = \widehat{\L}_S(\widehat{q}^*_S(q_{p'}; \omega); q_{p'}) = \widehat{\L}_S^*(q_{p'})&S\subseteq [n].
\end{align*}
This, along with \eqref{112} and \eqref{113}, tells us that $S^*$ is the estimated optimal model in the current step, and more exploration is needed.  

To see what will happen at $q_{p'}$, we consider two separate cases.
If $2q_{p'}\leq \widehat{q}^*_{S^*}(q_{p'}; \omega)$, then
\begin{align*}
T(\delta;\omega)<q_{p'+1} = 2q_{p'}\leq \widehat{q}^*_{S^*}(q_{p'}; \omega)\leq (1+\delta)q^*_{S^*},
\end{align*}
which implies 
\begin{align}
(1-\delta)q^*_{S^*}\stackrel{\eqref{113}}{\leq} q_{p'+1}\vee \widehat{q}^*_{S^*}(q_{p'+1}; \omega)\leq (1+\delta)q^*_{S^*}.\label{myku}
\end{align}
If $q_{p'}\leq \widehat{q}^*_{S^*}(q_{p'}; \omega)<2q_{p'}$, then
\begin{align*}
q_{p'+1} = \left\lceil\frac{q_{p'} + \widehat{q}^*_{S^*}(q_{p'}; \omega)}{2}\right\rceil\leq \widehat{q}^*_{S^*}(q_{p'}; \omega)\leq (1+\delta)q^*_{S^*},
\end{align*}
which also implies \eqref{myku}. 
But \eqref{myku} combined with \eqref{112} and \eqref{113} implies that $S^*$ is again the estimated optimal model in the $(p'+1)$th loop iteration. 
Applying the above argument inductively proves $S(B) = S^*$, i.e. \eqref{rr2}.
Note \eqref{myku} holds until the algorithm terminates, which combined with the termination criteria $q_{p(B)}\geq \widehat{q}^*_{S^*}(q_{p(B)}; \omega)\geq (1-\delta)q^*_{S^*}$ implies 
\begin{align}
1-\delta\leq\frac{q(B; \omega)}{q^*_{S^*}} = \frac{q_{p(B)}}{q^*_{S^*}}\leq 1+\delta.\label{261526}
\end{align}
In this case, \eqref{rr1} now follows by noting that $\delta$ can be set arbitrarily small. 

To exclude case (ii) for all sufficiently large $B$, we argue by contradiction. 
Suppose case (ii) occurs. In this situation, we consider the previous exploration rate before termination, $q_{p(B)}$, which, by definition, satisfies either condition in lines 9 or 16. Moreover, we denote by $q_{p(B)+1}$ the hypothetical next-step exploration rate assuming the collection process in either line 10 or line 19 were successful. By definition, 
\begin{align*}
q_{p(B)}\leq q(B; \omega)\leq q_{p(B)+1}\leq 2q_{p(B)}. 
\end{align*} 
Since $q(B; \omega)$ diverges as $B\to\infty$ as shown in step 1, so is $q_{p(B)}$. Therefore, by the same reasoning in case (i), $q_{p(B)+1}\leq (1+\delta)q^*_{S^*}$ for sufficiently large $B$. Consequently, the total exploration cost at $q(B; \omega)$ is upper bounded by 
\begin{align*}
(1+\delta)q^*_{S^*}\widehat{c}_\ex&\stackrel{\eqref{pj1}}{=}\frac{(1+\delta)B\widehat{c}_\ex}{c_\ex + \sqrt{\frac{c_\ex \gamma(S^*)}{k(S^*)}}}\stackrel{\eqref{1113}}{\leq} \frac{(1+\delta)^2 B c_\ex}{c_\ex + \sqrt{\frac{c_\ex \gamma(S^*)}{k(S^*)}}}\\
&< \left(1-\frac{1}{8}\sqrt{\frac{\gamma(S^*)}{c_\ex k(S^*)}}\right)B&\text{\bigg(since $\delta<\frac{1}{4}\sqrt{\frac{\gamma(S^*)}{c_\ex k(S^*)}}$\bigg)}\\
&< B - (n+1)c^*. 
\end{align*}
However, this implies that at $q(B; \omega)$, the budget is not running out, and one can draw at least one more exploration sample. Hence, we arrive at a contradiction. 

\subsection{Proof of \Cref{cvgh}}\label{sec:supp7}
\Cref{thm:AETC} shows that the exploitation budget $B_\ext = B-c_\ex q(B)$, as a function of $B$, diverges as $B\to\infty$ a.s.
Moreover, $S(B) = S^*$ for all sufficiently large $B$.
This implies that $\widehat{y}_{\widehat{S}^*}\to y_{S^*}$ and $\bt_{\widehat{S}^*}\to\bt_{S^*}$ a.s., proving the desired result.

\section{Proofs in \Cref{sec:lam}}\label{sec:supp4.3}

\subsection{Proof of \Cref{lemma:acv}}\label{sec:supp8}

To prove \Cref{lemma:acv}, we recall the notation in \eqref{haozi}: $\bar{\mu}_S = \sum_{i\in [q]}Q_{S,i}/q $ and $\bar{\mu}_0 = \sum_{i\in [q]}Q_{0,i}/q$. Explicit expressions for both $\widehat{a}_S$ and $\widehat{b}_S$ can be obtained using \eqref{lse}:  
\begin{align*}
&\widehat{b}_S = \left(\frac{1}{q}\sum_{i\in [q]}Q_{S, i}Q_{S, i}^\top-\bar{\mu}_S\bar{\mu}_S^\top\right)^{-1}\frac{1}{q}\sum_{i\in [q]}(Q_{S, i}-\bar{\mu}_S)Q_{0, i}& \widehat{a}_S= \bar{\mu}_0 - \bar{\mu}_S^\top\widehat{b}_S. 
\end{align*}
Consequently, 
\begin{align}
\lrmc &= \widehat{y}_S^\top\widehat{\beta}_S = \widehat{a}_S + \widehat{\mu}^\top_S \widehat{b}_S = \bar{\mu}_0- \widehat{b}_S^\top(\bar{\mu}_S- \widehat{\mu}^\top_S),\label{needc1}
\end{align}
which can be identified as \eqref{eq:ACVdefinition} with $\alpha = \widehat{b}_S$.

\subsection{Proof of \Cref{addback}}\label{sec:supp9}

We first note that $\lrmcstar$ is unbiased for $\mu_0$ and its variance is lower bounded by its CV counterpart (replacing $\widehat{\mu}_S$ by $\mu_S$) as $\V[\lrmcstar]\geq \sigma_S^2/q$ \cite[Eq. (2.3)]{han2023approximate}. Meanwhile, by the exploration-unbiasedness of $\widehat{\mu}_S$ and Cauchy--Schwarz inequality,  
\begin{align}
&\E[(\lrmcstar-\lrmc)^2\mid Z_\lfs]\nonumber\\
=&\ \E\left[\E[((b_S-\widehat{b}_S)^\top(\bar{\mu}_S-\widehat{\mu}_S))^2\mid \widehat{b}_S, Z_\lfs]\mid Z_\lfs\right]\nonumber\\
=&\ \E\left[\E[(b_S-\widehat{b}_S)^\top(\Cov[\widehat{\mu}_S\mid \widehat{b}_S, Z_\lfs] + \frac{1}{q}\widehat{\Sigma}_S)(b_S-\widehat{b}_S)\mid \widehat{b}_S, Z_\lfs]\mid Z_\lfs\right]\nonumber\\
\leq&\ \frac{\|\Sigma_S\|_2 + \|\widehat{\Sigma}_S\|_2}{q}\cdot\E[\|b_S-\widehat{b}_S\|_2^2 \mid Z_\lfs]\nonumber\\
\stackrel{\eqref{lo989}+\eqref{tdfg}}{\lesssim}&\ \frac{\sigma_S^2\|\Sigma_S\|_2}{q^2} = o(\V[\lrmcstar]),\label{very!!}
\end{align}
where the last inequality requires the linear model assumption~\eqref{lr}.  

\subsection{Proof of \Cref{khty}}\label{sec:supp10}

We first notice that the independent allocation samples consist of $\{Q_{\afs, i}\}_{i\in [q]}$ (exploration) and $\{W_{T, i}\}_{T\subseteq S, i\in [m_T]}$ (exploitation). For convenience, we assume $m_T\geq 1$ for every $T\subseteq S$ and $S = \{n-s+1, \ldots, n\}$ so that $\mu_{\afs} = (\mu_0, \mu_{[n]\setminus S}, \mu_{S})^\top$; the more general case can be made into such a form via reordering of indices. 

Recall $\bar{\mu}_S = \sum_{i\in [q]}Q_{S,i}/q$ in \eqref{haozi}. According to \eqref{eq:BLUE_estimator}, the MLBLUE $\widehat{\mu}_\blue$ is given by
\begin{align}
&\widehat{\mu}_\blue = \left(q\Sigma_{\afs}^{-1}+ \sum_{T\subseteq S}m_T\widetilde{R}_T^\top \Sigma_T^{-1}\widetilde{R}_T\right)^{-1}\left(q\Sigma_{\afs}^{-1}\bar{\mu}_\afs+\sum_{T\subseteq S}\widetilde{R}_T^\top {\Sigma}_T^{-1}\left(\sum_{\ell=1}^{m_T}W_{T,\ell}\right)\right)\nonumber\\
& = \left(I_{n+1}+ q^{-1}\Sigma_{\afs}\sum_{T\subseteq S}m_T\widetilde{R}_T^\top \Sigma_T^{-1}\widetilde{R}_T\right)^{-1}\left(\bar{\mu}_\afs+q^{-1}\Sigma_{\afs}\sum_{T\subseteq S}\widetilde{R}_T^\top {\Sigma}_T^{-1}\left(\sum_{\ell=1}^{m_T}W_{T,\ell}\right)\right),\label{loju}
\end{align}
where $\widetilde{R}_T =  (0, R_T)\in\{0,1\}^{|T|\times (n+1)}$ is restriction matrix mapping $\mu_\afs$ to $\mu_T$ with $ 0\in\R^{|T|\times (n-s+1)}$. 

To compute the first component of $\widehat{\mu}_\blue$, denote $\Sigma_{-S}$ as the covariance of $Q_{\afs\setminus S}$ and $\Sigma_{-S, S}$ as the covariance between $Q_{\afs\setminus S}$ and $Q_S$, and note
\begin{align*}
&
\Sigma_{\afs} = 
\begin{bmatrix}
\Sigma_{-S} & \Sigma_{-S, S}\\
\Sigma^\top_{-S, S} & \Sigma_S
\end{bmatrix}
&\sum_{T\subseteq S}m_T\widetilde{R}_T^\top \Sigma_T^{-1}\widetilde{R}_T = 
\begin{bmatrix}
0 & 0\\
0 & q^{-1}\Xi
\end{bmatrix}.
\end{align*}
It follows from the direct computation that 
\begin{align*}
I_{n+1}+ q^{-1}\Sigma_{\afs}\sum_{T\subseteq S}m_T\widetilde{R}_T^\top \Sigma_T^{-1}\widetilde{R}_T = 
\begin{bmatrix}
I_{n-s+1} & q^{-1}\Sigma_{-S, S}\Xi\\
0 & I_s + \Delta^{-1}
\end{bmatrix}.
\end{align*}
By the Schur complement, 
\begin{align*}
\left(I_{n+1}+ q^{-1}\Sigma_{\afs}\sum_{T\subseteq S}m_T\widetilde{R}_T^\top \Sigma_T^{-1}\widetilde{R}_T\right)^{-1} &= \begin{bmatrix}
I_{n-s+1} & -q^{-1}\Sigma_{-S, S}\Xi(I_s+\Delta^{-1})^{-1}\\
0& (I_s + \Delta^{-1})^{-1}
\end{bmatrix}\\
&= \begin{bmatrix}
I_{n-s+1} & -\Sigma_{-S, S}\Sigma_S^{-1}(I_s+\Delta)^{-1}\\
0& (I_s + \Delta^{-1})^{-1}
\end{bmatrix},
\end{align*}
from which we can read the first row of $(I_{n+1}+ q^{-1}\Sigma_{\afs}\sum_{T\subseteq S}m_T\widetilde{R}_T^\top \Sigma_T^{-1}\widetilde{R}_T)^{-1}$ as 
\begin{align}
&\left(1, 0, \cdots, 0, -\Sigma_{Q_0, S}\Sigma_S^{-1}(I_s+\Delta)^{-1}\right)\in\R^{n+1}&\Sigma_{Q_0, S} = \Cov[Q_0, Q_S]. \label{hua1}
\end{align}
We also compute 
\begin{align}
&q^{-1}\Sigma_{\afs}\sum_{T\subseteq S}\widetilde{R}_T^\top {\Sigma}_T^{-1}\left(\sum_{\ell=1}^{m_T}W_{T,\ell}\right) = 
\begin{bmatrix}
q^{-1}\Sigma_{-S,S}\chi\\
q^{-1}\Sigma_S\chi
\end{bmatrix}&
\chi = \sum_{T\subseteq S}R_T^\top {\Sigma}_T^{-1}\left(\sum_{\ell=1}^{m_T}W_{T,\ell}\right).\label{hua2}
\end{align}
Substituting \eqref{hua1} and \eqref{hua2} into \eqref{loju} yields that
\begin{align}
\widehat{\mu}_{0,\blue} &= \bar{\mu}_0 - \Sigma_{Q_0, S}\Sigma_S^{-1}(I_s+\Delta)^{-1}\bar{\mu}_S+ \frac{1}{q}\Sigma_{Q_0,S}\chi-\frac{1}{q}\Sigma_{Q_0, S}\Sigma_S^{-1}(I_s+\Delta)^{-1}\Sigma_S\chi\label{shaoxin1}\\
& = \bar{\mu}_0 - b_S^\top(I_s+\Delta)^{-1}\bar{\mu}_S+ \frac{1}{q}\Sigma_{Q_0,S}\chi-\frac{1}{q}b_S^\top(I_s+\Delta)^{-1}\Sigma_S\chi,\nonumber
\end{align}
where the last step follows from the fact $b_S = \Sigma_S^{-1}\Sigma_{Q_0, S}^\top$. 
Assuming $\|\Delta\|_2<1$, $(I_s+\Delta)^{-1} = \sum_{k=0}^\infty(-\Delta)^{k}$, 
which is plugged into \eqref{shaoxin1} to yield 
\begin{align*}
\widehat{\mu}_{0,\blue} &=  \bar{\mu}_0 -b_S^\top\sum_{k=0}^\infty(-\Delta)^{k}\bar{\mu}_S+ \frac{1}{q}\Sigma_{Q_0,S}\chi-\frac{1}{q}b^\top_S\sum_{k=0}^\infty(-\Delta)^{k}\Sigma_S\chi\\
& = \bar{\mu}_0 -b_S^\top\sum_{k=0}^\infty(-\Delta)^{k}\bar{\mu}_S-\frac{1}{q}b^\top_S\sum_{k=1}^\infty(-\Delta)^{k}\Sigma_S\chi\\
& = \bar{\mu}_0 -b_S^\top\sum_{k=0}^\infty(-\Delta)^{k}\bar{\mu}_S+b_S^\top\sum_{k=0}^\infty(-\Delta)^{k}\widehat{\mu}_S &\text{$(\widehat{\mu}_S=\Xi^{-1}\chi)$}\\
& = \lrmcstar - b_S^\top\sum_{k=1}^\infty(-\Delta)^k(\bar{\mu}_S-\widehat{\mu}_S)\\
& = \lrmcstar + b_S^\top(I_s+\Delta)^{-1}\Delta(\bar{\mu}_S-\widehat{\mu}_S)\\
& = \lrmcstar + \mathcal O(\|\Delta\|_2\|\bar{\mu}_S - \widehat{\mu}_S\|_2)&\text{($\|\Delta\|_2\leq 1/2$)}
\end{align*}

\subsection{Proof of \Cref{ajan}}\label{sec:supp11}

Suppose the optimal allocation achieving $\mse_S^*$ contains $q$ samples involving the high-fidelity model. According to \cite[Lemma 3.5]{Schaden_2020}, the MSE/variance of the corresponding MLBLUE is nonincreasing if we augment each sample containing the high-fidelity model to a full sample containing all low-fidelity models. This procedure comes with an additional cost bounded by $q(c_\ex-c_0)$. Consequently, the resulting estimator using the augmented samples has its allocation in $\mathcal{P}(S)\cup\{\afs\}$. By the definition of $\mse_S(B)$ and its linear scaling in $1/B$, 
\begin{align*}
\mse_S(B)&= \frac{B+q(c_\ex-c_0)}{B}\mse_S(B+q(c_\ex-c_0))\\
&\leq \frac{B+q(c_\ex-c_0)}{B}\mse_S^*(B)\\
&\leq \frac{qc_0+q(c_\ex-c_0)}{qc_0}\mse_S^*(B)&\text{($qc_0\leq B$)}\\
&=\frac{c_\ex}{c_0}\cdot\mse_S^*(B).
\end{align*}

\section{Additional details for numerical setup}\label{sec:supp12}

\begin{figure}[htbp]
\begin{center}
\includegraphics[width=0.45\textwidth]{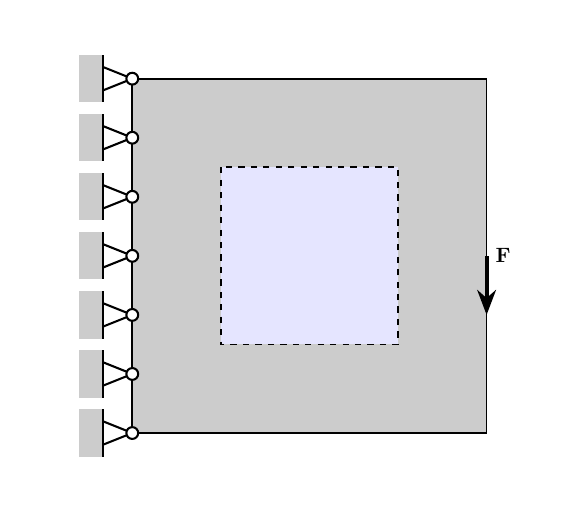}
\caption{\small Geometry, boundary conditions, and loading for a linear elastic structure with the square domain.}\label{fig:struct}
  \end{center}
\end{figure}

We provide additional details for the parametric PDE example in \Cref{sec:elastic-pde}. The PDE is described as 
\begin{align*}
  \nabla \cdot \bm{\sigma}(\bm{p}) = -\bm{F},
\end{align*}
where $\bm{\sigma} \in \R^{2 \times 2}$ is the two-dimensional plane stress tensor, related to the displacement $u$ through
\begin{align}\label{elliptic_pde_stochastic}
  \bm{\sigma} &= \left(\begin{array}{cc} \sigma_x & \sigma_{x y} \\ \sigma_{x y} & \sigma_y \end{array}\right), & 
    \left(\begin{array}{c} \sigma_x \\ \sigma_y \\\sigma_{x y} \end{array}\right) = \frac{\kappa(\bm{p},\bm{x})}{1 - \nu^2} \left(\begin{array}{ccc} 1 & \nu & 0 \\ \nu & 1 & 0 \\ 0 & 0 & 1-\nu \end{array}\right) \left(\begin{array}{c} u_x \\ u_y \\ \frac{1}{2}\left(u_x + u_y\right) \end{array}\right),
\end{align}
where $\kappa$ is the Young's modulus defined in \eqref{eq:kappa-def}, and $\nu$ is Poisson's ratio that we set to $\nu = 0.3$. 
The PDE was defined over a square spatial domain $D = [0,1]^2$.
The exact geometry, boundary conditions, and loading used are shown in \Cref{fig:struct}.
Randomness was introduced via the parameters $\bm p \in \mathcal{P} \subset \R^4$ which was a random vector with independent components uniformly distributed on $[-1,1]$ that perturbed a truncated Karhunen-Lo\`{e}ve expansion of the Young's modulus
\begin{align}\label{eq:kappa-def}
  \kappa(\bm p, \bm x) = 1 + 0.5 \sum_{i=1}^4 \sqrt{\lambda_i} \phi_i(\bm x) p_i,
\end{align}
where $(\lambda_i, \phi_i)$ are ordered eigenpairs of an exponential covariance kernel on $D$. (See \cite[Section 7.1]{xu2022bandit} for more details.) 

In this study, we construct a multi-fidelity hierarchical model ensemble by coarsening the mesh parameter $h$ of a finite element solver with standard bilinear square isotropic finite elements defined on a rectangular mesh \cite{andreassen2011efficient}. Starting with the highest fidelity model, with mesh size $h =2^{-7}$ as the high-fidelity model, we create four low-fidelity models based on computationally cheaper but less accurate discretizations: $h = 2^{-4}, 2^{-3}, 2^{-2}$, and $2^{-1}$. We then use this model ensemble to compute the mean of our scalar QoI, which is the structural \emph{compliance}, i.e., the energy norm of the displacement solution $\bm{u} = (u_x, u_y)^\top$, 
\begin{align*}
  E(\bm{p}) = \int_D (\bm{u}(\bm{x},\bm{p}) \cdot \bm{F}) \mathrm{d}\bm{x}.
\end{align*}
where the loading force $\bm{F}$ is non-zero only at the lower right of the structure, having value $\bm{F} = (0, -1)^\top$ there. 

We assume that the computational cost of each model is deterministic and inversely proportional to the mesh size squared ($h^2$); this corresponds to using a linear solver of optimal linear complexity. Moreover, we normalize cost so that the lowest fidelity model has unit cost, i.e., $c_0 = 4096, c_1 = 64, c_2 = 16, c_3 = 4, c_4 = 1$. The correlations between the outputs of $Q_0$ and $Q_1, Q_2, Q_3, Q_4$ are $0.976$, $0.940$, $0.841$, $-0.146$, respectively. 
\end{appendix}

\end{document}